\documentclass[showpacs,prb,aps,superscriptaddress,twocolumn,floatfix]{revtex4}
\usepackage[T1]{fontenc}
\usepackage{graphicx}
\usepackage{amsmath,float}
\usepackage{amsfonts} 
\usepackage{amsthm} 
\usepackage{amssymb} 
\usepackage{multirow}
\usepackage[a4paper,left=1cm,right=1cm,top=2cm,bottom=2cm]{geometry}
\usepackage{bbm}
\newcommand{\K}{\vec{K}}
\newcommand{\Kp}{\vec{K}'}
\newcommand{\kk}{\vec{k}}
\newcommand{\q}{\vec{q}}
\newcommand{\rr}{\vec{r}}
\newcommand{\ep}{\epsilon}
\newcommand{\be}{\begin{eqnarray}}
\newcommand{\ee}{\end{eqnarray}}
\newcommand{\Ai}{\mbox{Ai}}
\newcommand{\Bi}{\mbox{Bi}}

\begin{document}
\title{WKB analysis of edge states in graphene in a strong magnetic field.}

\author{Pierre Delplace}
\affiliation{Laboratoire de Physique des Solides, UMR 8502 du CNRS,
              Universit\'e Paris-Sud 11, 91405 Orsay, France.}

\author{Gilles Montambaux}
\affiliation{Laboratoire de Physique des Solides, UMR 8502 du CNRS,
              Universit\'e Paris-Sud 11, 91405 Orsay, France.}

\today
\begin{abstract}We investigate the fine structure of the  edge states energy spectrum for zigzag and armchair ribbons of graphene in a strong magnetic field. At low energy, the spectra can be described by an effective Schr\"odinger Hamiltonian with a double well potential, symmetric in the zigzag case and asymmetric in the armchair case. We develop a semiclassical formalism based on the WKB approximation to calculate analytically the energy spectrum for the two types of edges, including regions which were not studied earlier. Our results are in very good quantitative agreement with  numerical calculations. This approach leads to a qualitative description of the spectra in terms of the quantization of unusual classical orbits in the real space.
\end{abstract}

\maketitle

																									 \section{Introduction}

The fascinating properties of graphene originate from its perfect two-dimensional character and from its unusual electronic structure where valence and conduction bands touch at two inequivalent points of reciprocal space. In the vicinity of these two points, the dispersion relation is linear and charge carriers behave as massless Dirac fermions\cite{wallace} with striking consequences on the electronic properties, like the unusual 2D quantum Hall effect.\cite{geim,kim} Recent studies have also addressed electronic properties of confined graphene structures like dots, rings or nanoribbons. In particular, nanoribbons have been suggested as potential candidates for replacing electronic components in future nanoelectronic and nanospintronic devices.\cite{novoselov,louie} In that perspective, the role of edge states is essential. In the absence of magnetic field, edge states can emerge for particular types of edges.\cite{fujita,nakada} Such localized electronic states can  be described analytically at low energy by using the Dirac equation with the appropriate boundary conditions.\cite{bf1} More recent theoretical works have investigated the metallic nature of such edge states by considering staggered on-site potentials,\cite{yao} surface perturbations,\cite{klos} spin-orbit coupling,\cite{li} edge dopping,\cite{xu} and sophisticated terminations of the honeycomb lattice.\cite{akhmerov,ihnatsenka,baranger} Experimental efforts have been performed to observe different natures of edges\cite{kobayashi,chuvilin,koskinsen} and study charge transport in ribbons.\cite{han,gallagher}    

This paper deals with the structure of edge states in a strong magnetic field, a problem which has been the subject of recent interest. Although this problem is somehow reminiscent of the case of usual 2D electrons,\cite{halperin,büttiker} it is much richer since here charge carriers are massless particules and the structure of the edge states depends  of the nature of the edges.  In particular, Brey and Fertig first calculated within the tight-binding model  the band structures of ribbons in a magnetic field with two types of boundary conditions: the so-called zigzag and armchair edges.\cite{bf} In addition, they have shown that these two types of edges can be described at low energy by a Schr\"odinger Hamiltonian with a harmonic potential. In the zigzag case, this potential is similar to the one of the usual integer quantum Hall problem,\cite{halperin} whereas in the armchair case, the potential presents a specific asymmetry. Moreover, these edge states have been recently investigated numerically  in graphene rings,\cite{bahamon} and in graphene ribbons by considering an anisotropic hopping parameter,\cite{dahal} or by taking account for a possible quantum Hall ferromagnetism.\cite{gusynin}

However, the energy dependence of the Landau levels when approching the edges has only been discussed briefly. The goal of this paper is to investigate in details the fine structure of the edge states energy spectrum for zigzag and armchair ribbons  and provide a complete \emph{quantitative analytical} description  within the WKB approximation. The paper is organized as follows: In section \ref{tbs}, we recall and discuss in details the tight-binding spectra for zigzag and armchair ribbons with and without magnetic field. When a field is applied, we point out peculiar features of the edge states   which we describe analytically in the rest of the paper. Following the procedure suggested by Brey and Fertig,\cite{bf}   we derive in section \ref{asd} a simple effective Hamiltonian (essentially the squared Dirac Hamiltonian) with a potential depending on the boundary conditions. Then, in section \ref{sec:sc}, we present a semiclassical framework to calculate analytically the low energy spectra. Two methods are used: the first one is based on the Bohr-Sommerfeld quantization of the action and leads to a simple qualitative picture in terms of classical skipping orbits in section \ref{secarea}. However, this method does not properly describe the case where the classical cyclotron radius is of the order of the distance  to the edge. Therefore, we develop a more sophisticated semiclassical approach, based on the WKB approximation, that leads to analytical quantitative results in agreement with numerical calculations. The details of this approach are given in the appendix. We conclude in section \ref{sec:conclu}. 


																				  		 \section{Tight-binding spectrum}
																				  		\label{tbs}
																							\subsection{Zero magnetic field}

\begin{figure}[!ht]
	\centering
		\includegraphics[width=9cm]{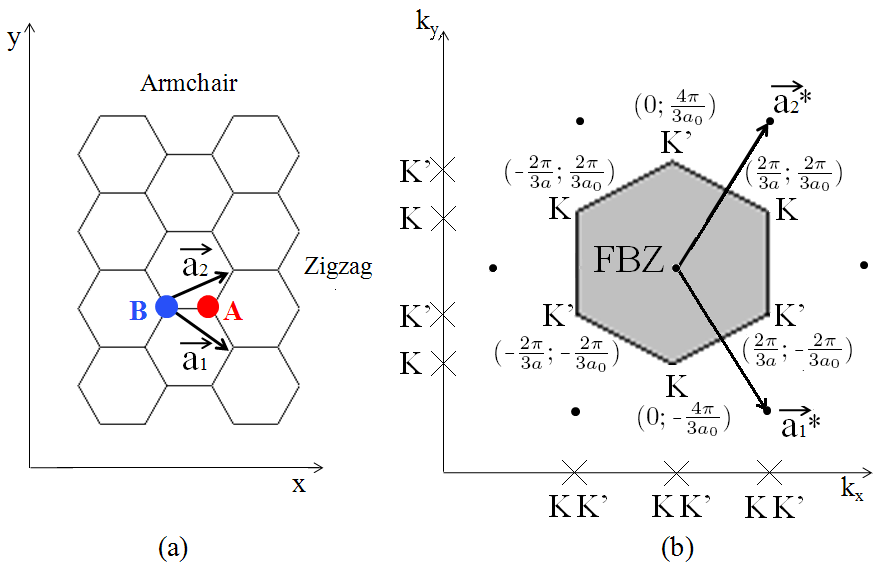}
	\caption{(a) Sheet of graphene with zigzag and armchair edges. (b) The black dots represent the nodes of the triangular reciprocal  lattice whose $\vec{a}_1^{*}$ and $\vec{a}_2^{*}$ are the basis vectors. The Dirac points $\K$ and $\Kp$ are located at  the corners of the FBZ.  The projections of the $\K$ and $\Kp$ points on the  $k_x$ and $k_y$ axes  have been represented by crosses.}
	\label{fig:rotation}
\end{figure}

\begin{figure}[!ht]
	\centering
		\includegraphics[width=9cm]{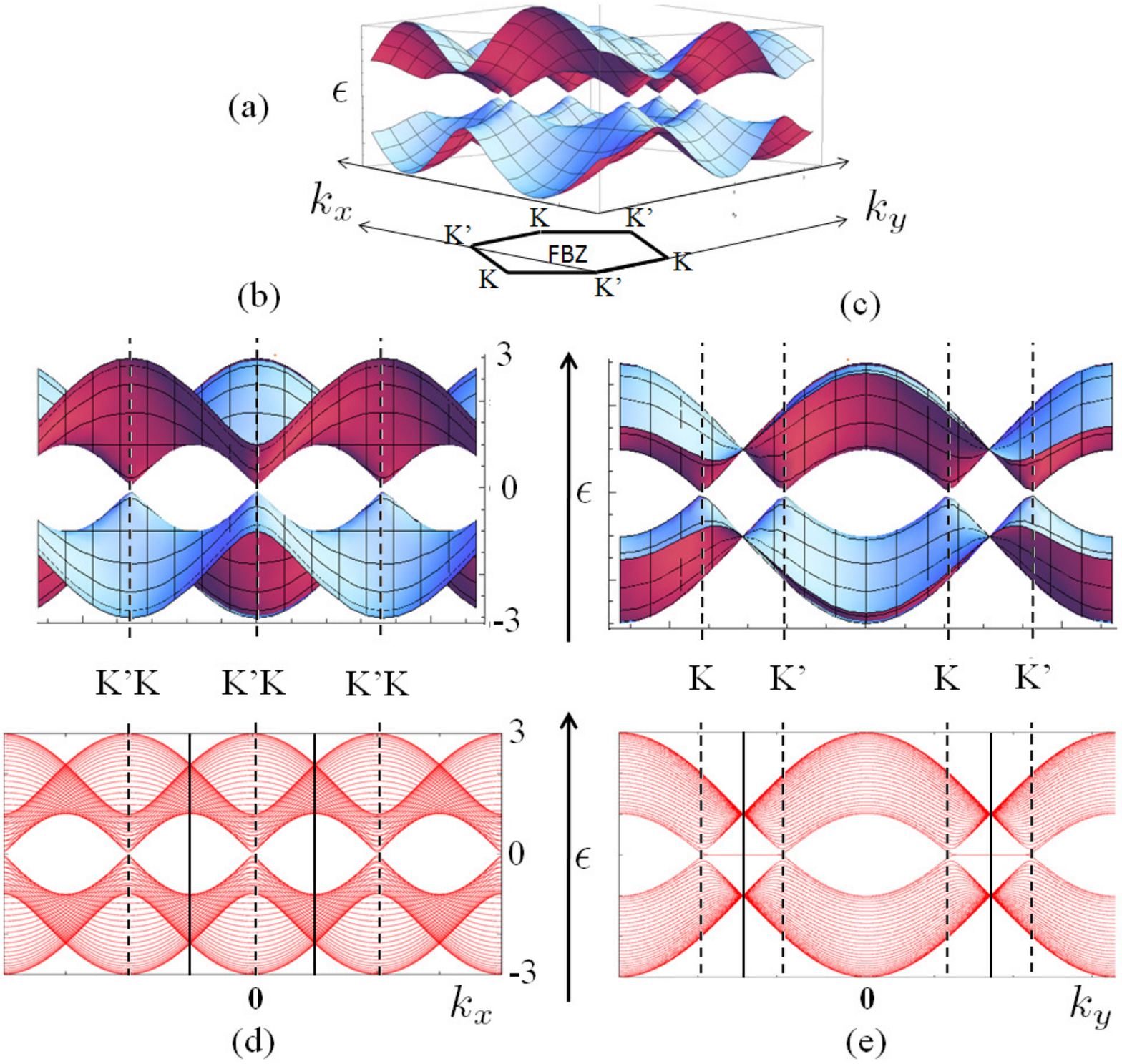}
	\caption{(a) Energy spectrum $\ep(\vec{k})$ of an infinite sheet of graphene.  Projection of this   spectrum (b) along the $k_y$ axis onto the plane $(\ep,k_x)$ and (c) along the $k_x$ axis onto the plane $(\ep,k_y)$.  Band structure for an infinite armchair ribbon (d),   and  for a zigzag ribbon (e). Their widths are $L=49a_0/2$ for the armchair ribbon and $L=49\sqrt{3}a_0/2$ for the zigzag ribbon. The positions of the $\K$ and $\Kp$ points are represented by dashed vertical lines and the 1D FBZ is delimited by continuous vertical lines. The energies are given in units of the hopping parameter $t$.}
		\label{fig:projection}
\end{figure}

We  briefly recall  the band structure of infinite graphene  ribbons within the tight-binding picture. For that purpose, we first consider the  Hamiltonian $H$ of  an infinite 2D  sheet of graphene, and we use the Bloch theorem in both $x$ and $y$ directions. As the honeycomb lattice has two carbon atoms per cell (that we call A and B, see Fig. \ref{fig:rotation}), the Bloch wave function is written as~:
\begin{eqnarray}
\left|\Psi_{\kk}\right\rangle=\sum_j{e^{i\kk {\vec R}_j}\left[\Psi_A \left|A_j\right\rangle+\Psi_B\left|B_j\right\rangle\right]}
\end{eqnarray}
where $\vec{R}_j$ are the vectors of the triangular Bravais lattice. Then, within the tight-binding model, the Hamiltonian is written as~:
\begin{equation}
H=t \left(
\begin{array}{cc}
0&f(\kk)\\
f^{*}(\kk)&0\\
\end{array}
\right)
\label{tb}
\end{equation}
in the basis of the two sublattices $(\Psi_A,\Psi_B)$ where $t$ is the hopping parameter between nearest neighbors and with~:
\begin{eqnarray}
f(\kk)=-1-\exp{(i\kk \cdot\vec{a}_1 )}-\exp{(i\kk \cdot\vec{a}_2 )}
\end{eqnarray}
%
where $\vec{a}_1$ and $\vec{a}_2$ are the basis vectors of the triangular  Bravais lattice with $\left\|\vec{a}_1\right\|=\left\|\vec{a}_2\right\|\equiv a_0$. The lattice parameter $a_0$ is related to the carbon-carbon distance $a=0.142$nm by $a_0=\sqrt{3}a$. The dispersion relation  given by $\epsilon(\kk)=\pm t \left|f(\kk)\right|$ (see Fig. \ref{fig:projection}(a)) consists in two bands which touch at the corners of the first Brillouin zone (FBZ).  The positions of these points are given by the condition $f(\vec{\textit{K}})=0$, that is~:
\begin{eqnarray*}
\vec{K}=\frac{1}{3}(\vec{a}_1^{*}-\vec{a}_2^{*})\ \ \ \ \
\vec{K}'=-\frac{1}{3}(\vec{a}_1^{*}-\vec{a}_2^{*})
\end{eqnarray*}
where $\vec{a}_1^{*}$ and $\vec{a}_2^{*}$ are the basis vectors of the reciprocal lattice. The spectrum is linear near these points.
The choices of the unit cell and of   the axes are illustrated in  Fig. \ref{fig:rotation}(a). The positions of the so-called $\K$ and $\Kp$ Dirac points are shown in Fig. \ref{fig:rotation}(b), where we have also indicated their projections onto the axis $k_x$ and $k_y$.

We consider now infinite ribbons with either zigzag or armchair edges. In these finite geometries, the Bloch theorem can be used only along the infinite direction, and the boundary condition along the finite direction yields a finite number of  bands. These bands,  computed in the tight-binding model, are displayed in Figs. \ref{fig:projection} (d-e).  It is seen that the spectra of these ribbons correspond to the projection of the 2D spectrum along the $k_x$ axis (armchair edge) or the $k_y$ axis (zigzag edge). The projected points $\K$ and $\K'$ coincide in the armchair case but they do not in the zigzag case.  This remark illustrates the fact that for the armchair case, the two valleys are admixed by the boundary condition as we will discuss later.  Finally, note that  the flat level between the two points $\K$ and $\K'$ for zigzag edge is not captured by the projection of the bulk result. This zero energy state  is  localized near the edges  and its existence depends on the boundary conditions.\cite{nakada,bf1,hatsugai} This edge state exists whitout any magnetic field, contrary to the edge states in the quantum Hall regime which are studied in this paper.

\begin{figure}[!ht]
	\centering
		\includegraphics[width=9cm]{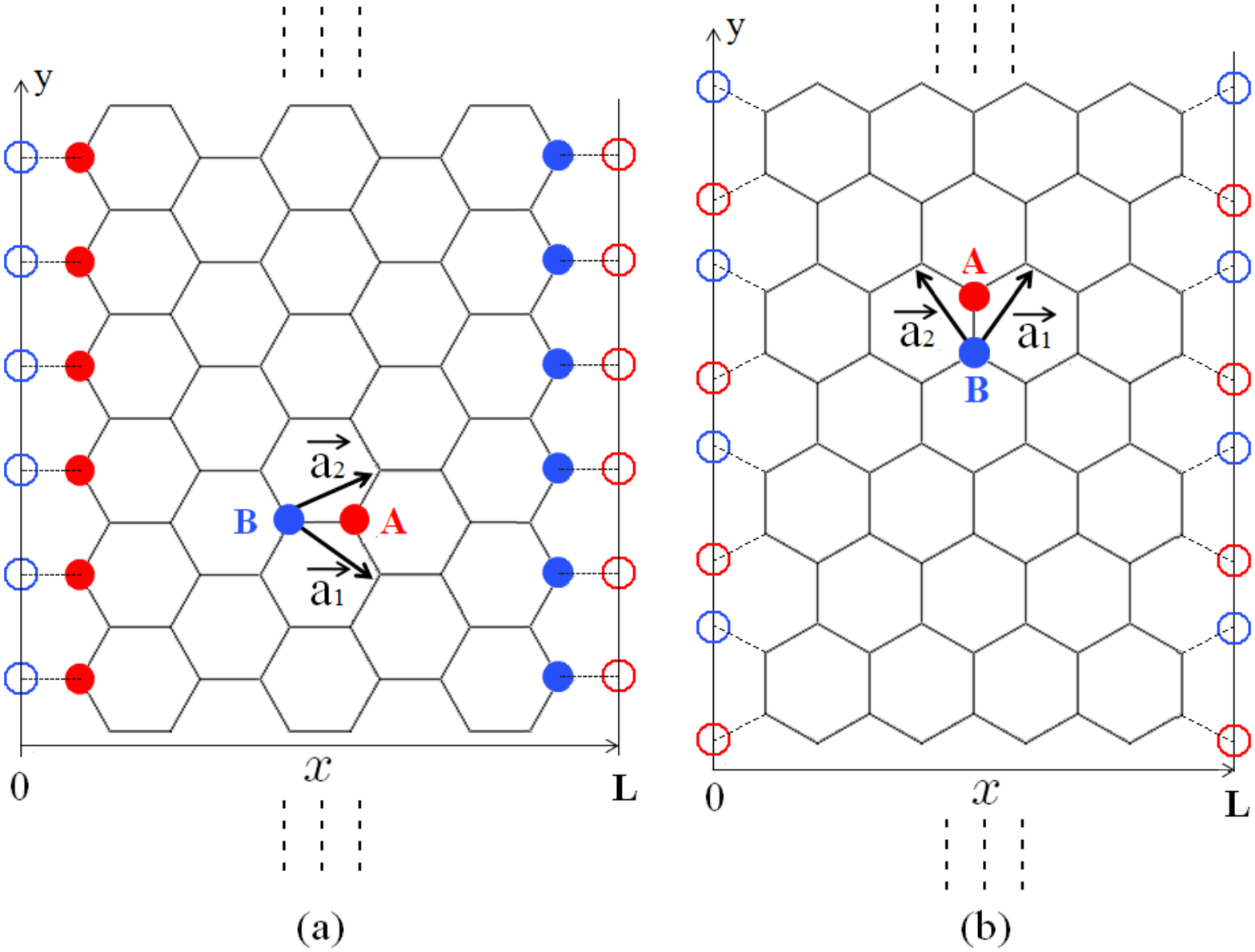}
	\caption{Graphene ribbons of width L with (a) zigzag edges and (b) armchair edges. We consider the $y$ direction as infinite. The circles represent empty sites and define the edges.}
		\label{fig:rubans}
\end{figure}

From now on, we define the geometry of the armchair and zigzag ribbons as shown in Fig. \ref{fig:rubans}~: the ribbon is infinite along the $y$ direction and it has a finite width $L$ in the $x$ direction. Within this convention, we have $K_y\neq K'_y$ for the zigzag ribbons and $K_y=K'_y$ for the armchair ones.

																					\subsection{Perpendicular magnetic field}
																				 \label{pmf}
\begin{figure}
	\centering
		\includegraphics[width=9cm]{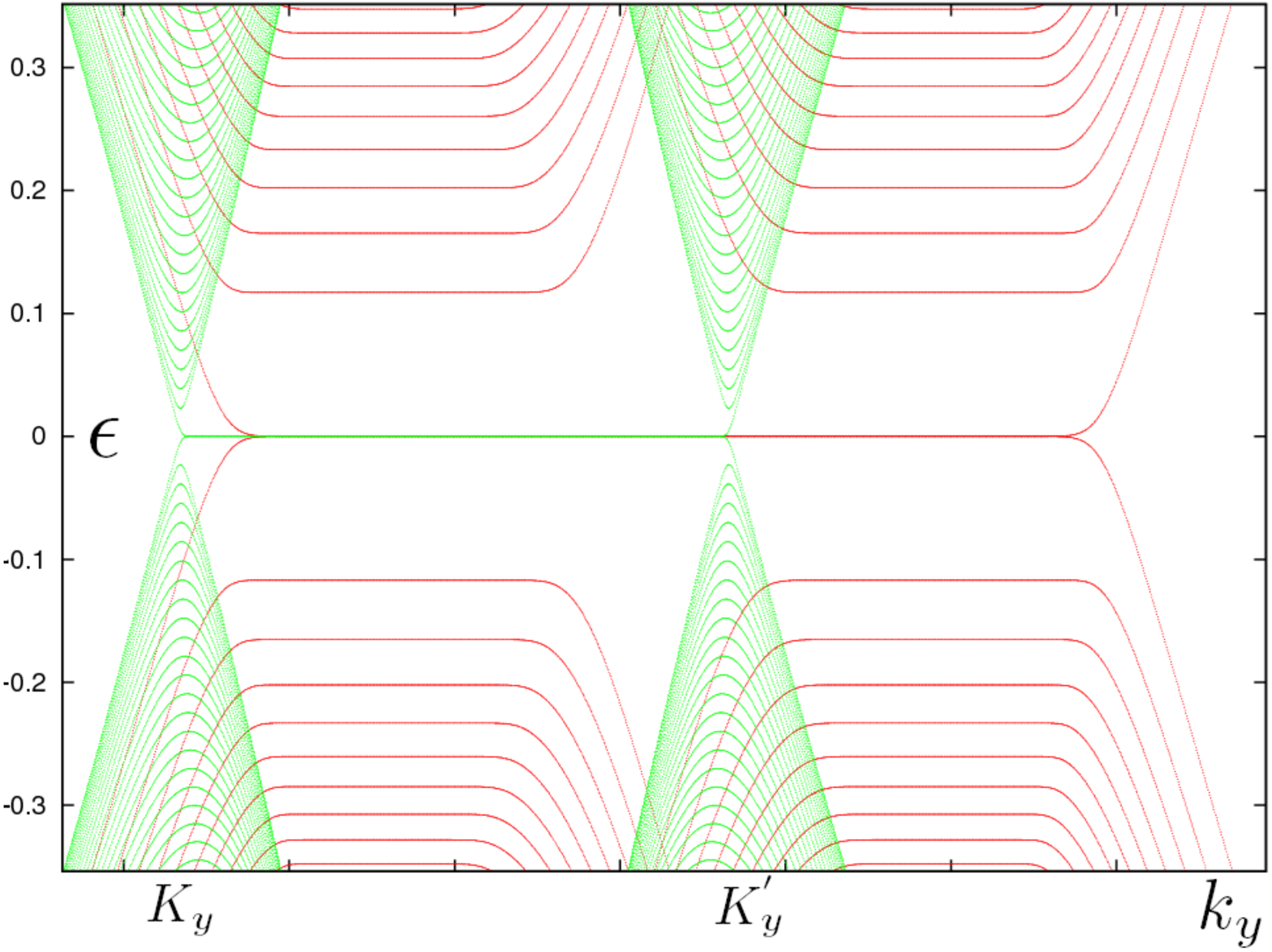}
	\caption{Tight-binding  spectrum at low energy for a zigzag ribbon with (red) and without (green) magnetic field. The dimensionless magnetic flux is $\phi=0.00126\phi_0$ and the width is $L=199 \sqrt{3}a_0/2 $.}
		\label{fig:champzz}
\end{figure}

\begin{figure}
	\centering
		\includegraphics[width=9cm]{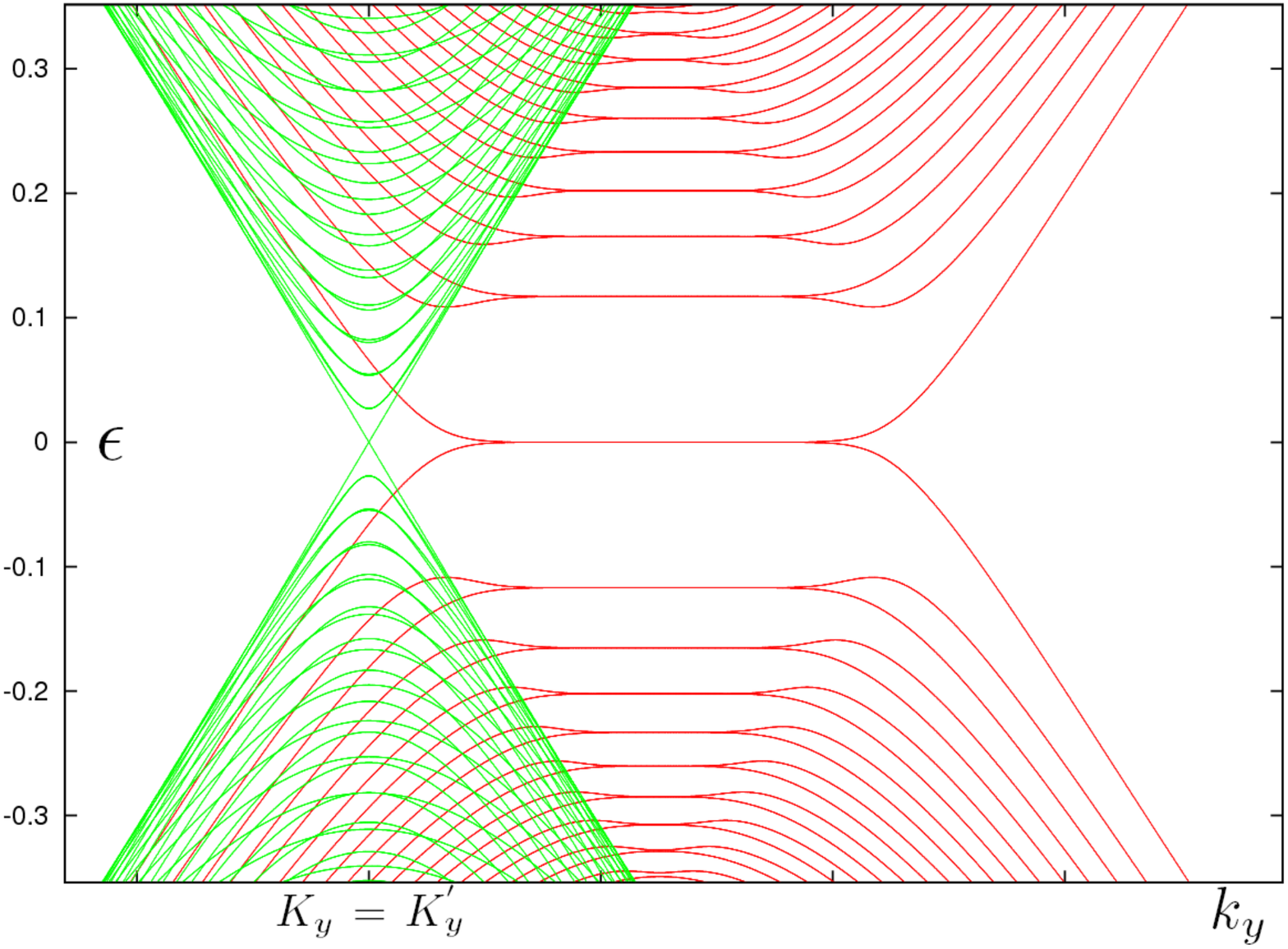}
	\caption{Tight-binding   spectrum at low energy for an armchair ribbon with (red) and without (green) magnetic field. The dimensionless magnetic flux is $\phi=0.00126\phi_0$ and the width is $L=199a_0/2$.}
		\label{fig:champarm}
\end{figure}

We now apply a perpendicular magnetic field $\vec{B}=-B\vec{e}_z$ on each type of ribbon and calculate numerically the band structure within the tight-binding model.\cite{bf} The finite width of the ribbons along the $x$ axis leads us to work with the Landau gauge $A_y = -Bx$. Because of the vector potential $A_y$, the hopping parameter $t$ (taken as unity in the rest of the paper), takes a Aharonov-Bohm phase $t\rightarrow t e^{i\frac{2\pi}{\Phi_0}\int d\vec{l}\cdot \vec{A}}$, where $\phi_0=h/e$ is the magnetic flux quantum. We now study the low energy spectrum for a given magnetic flux $\phi$ through an elementary plaquette. In Fig. \ref{fig:champzz}, we show the  low energy tight-binding spectra for the zigzag ribbon with and without magnetic field in the same plot. The spectra for the armchair case are shown in Fig. \ref{fig:champarm}. The zero field cones are progressively transformed into flat Landau levels with the expected $\epsilon_n = \pm t\sqrt{2 \pi \sqrt{3} \ n\ \phi/\phi_0 }$ behaviour.\cite{maclure}

\begin{figure}[!ht]
	\centering
		\includegraphics[width=9cm]{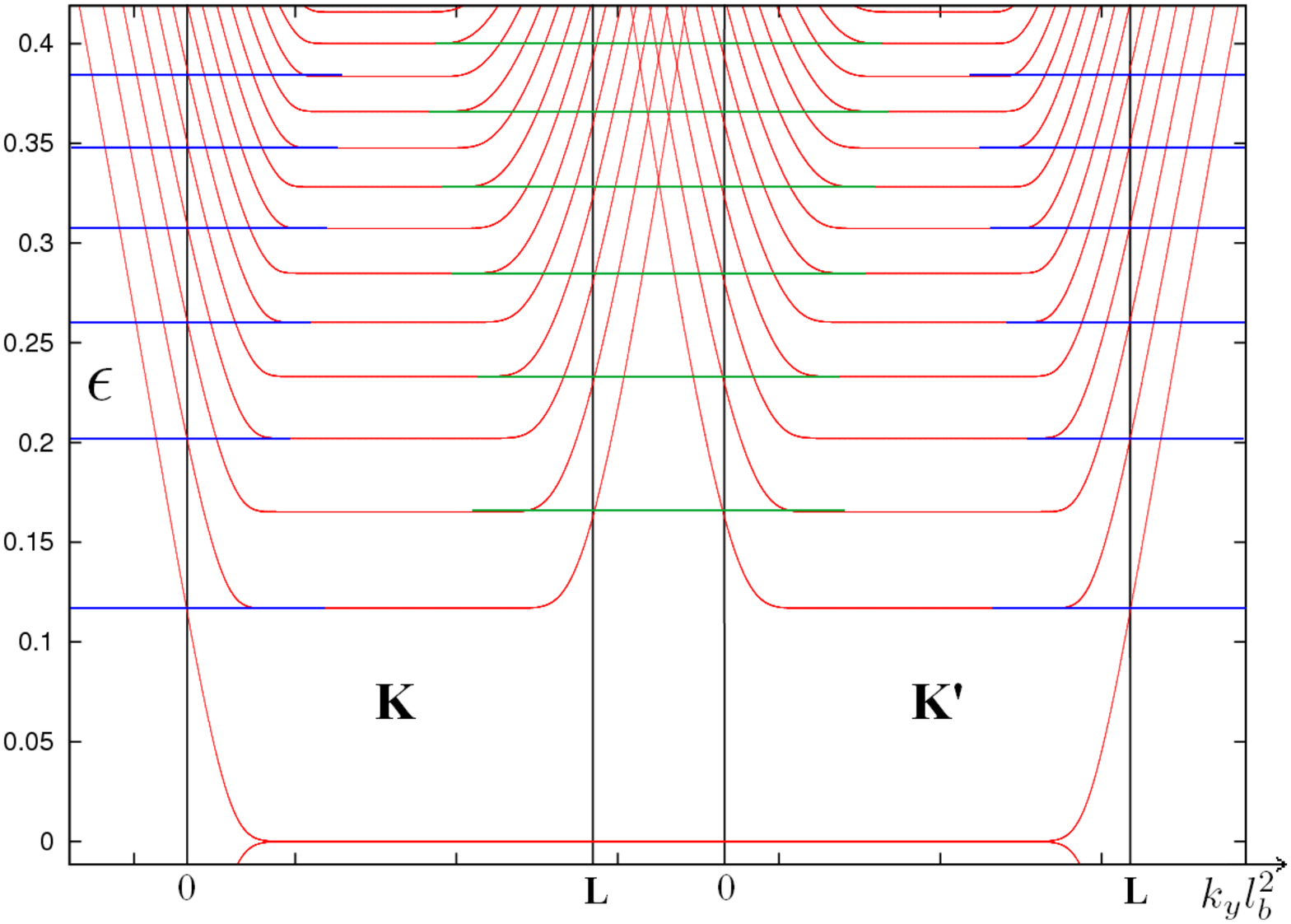}
	\caption{Tight-binding low energy levels in the quantum Hall regime as a function of the position $k_y \ell_B^2$ along the $x$ axis,  for a zigzag  ribbon. The edges are represented by vertical black lines. Green and blue horizontal lines indicate the position of the bulk Landau levels  for comparison with the position of the levels at the edges.  Note  the mirror symmetry of the spectrum between the two valleys $\K$ and $\Kp$. The energy is given in units of $t$.}
		\label{fig:zzx}
\end{figure}

\begin{figure}[!ht]
	\centering
		\includegraphics[width=9cm]{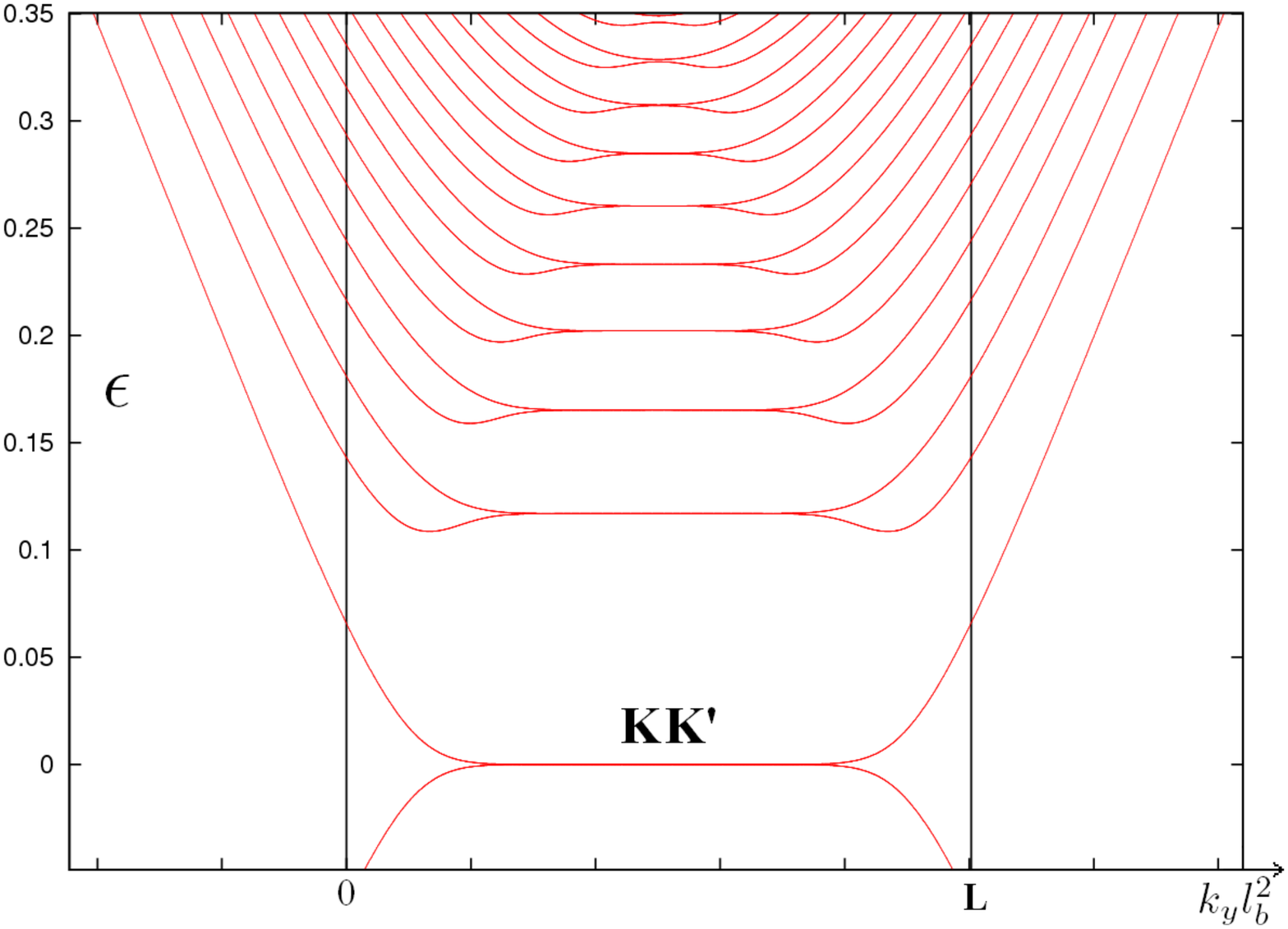}
	\caption{Tight-binding low energy levels in the quantum Hall regime as a function of the position $k_y \ell_B^2$,  for an armchair  ribbon. The edges are represented by vertical black lines. The energy is given in units of $t$.}
		\label{fig:halparm}
\end{figure}

We now comment on several important features of these spectra which to our knowledge have not been discussed in the literature. Since in the chosen Landau gauge, a state $k_y$   is centered at the position  $k_y \ell_B^2$ along the $x$ direction, where $\ell_B=\sqrt{\hbar/e B}$ is the  magnetic length, the variation of the energy levels can be interpreted as  a function of this position, as redrawn in Figs. \ref{fig:zzx} and \ref{fig:halparm}.  In these figures,  we have indicated the position of the edges along the $x$ direction, at low energy. Indeed, the position of one edge is fixed by the position of the  Dirac points in zero field and the other edge is located at  a distance $\Delta q_y =L/\ell_B^2$ where $L$ is the width of the ribbon and where $q_y=k_y-K^{(\prime)}_y$. As the two valleys are not admixed in the zigzag case, this operation must be performed for both $\K$ and $\Kp$ valleys, as seen in Fig. \ref{fig:zzx}. Therefore, it is clear that the dispersive character of the levels corresponds to the edge states, as first discussed by Halperin in the context of the integer quantum Hall effect of massive particules in  two dimensional gases.\cite{halperin} However, the band structure for graphene is more complex and  depends on the type of edge as we discuss now.
For the zigzag edge, we first emphasize that, inside a given valley, \textit{the spectrum is not symmetric}:  the energy levels are not identical on left and right sides. Moreover,  exactly at the left   and   right edges, the  energy levels $\epsilon^{edge}$ take peculiar values~: they alternatively take the value of higher  bulk Landau levels $\epsilon^{bulk}$.  This correspondence is displayed by horizontal color lines in Fig. \ref{fig:zzx}. More precisely, for the $\K$ valley, we have the relation $\epsilon_n^{edge}=\epsilon_{2n+1}^{bulk}$ with $n\geq 0$ (blue lines) on the left edge  whereas we have $\epsilon_n^{edge}=\epsilon_{2n}^{bulk}$  (green lines) with $n>0$ for the right edge. In addition, we have the same structure in the $\Kp$ valley, provided the role of the two edges is permuted. This remarkable distribution of the edge states will be explained in section \ref{sec:results}.
The armchair case, in Fig. \ref{fig:halparm}, is quite different. Each Landau level is  doubly degenerate, a direct consequence of the valley admixing. The degeneracy lifting takes place close to the edge with  a non monotonous behaviour for one every two levels, that  will be explained in section \ref{sec:results}.  
In the following we present a qualitative and quantitative description of these spectra within a semiclassical analytical approach.

																							\section{Low energy effective Hamiltonian}
																							\label{asd}
																							\subsection{The bulk Hamiltonian}
                                               \label{secbh}
 																
\begin{figure}[!ht]
	\centering
		\includegraphics[width=9cm]{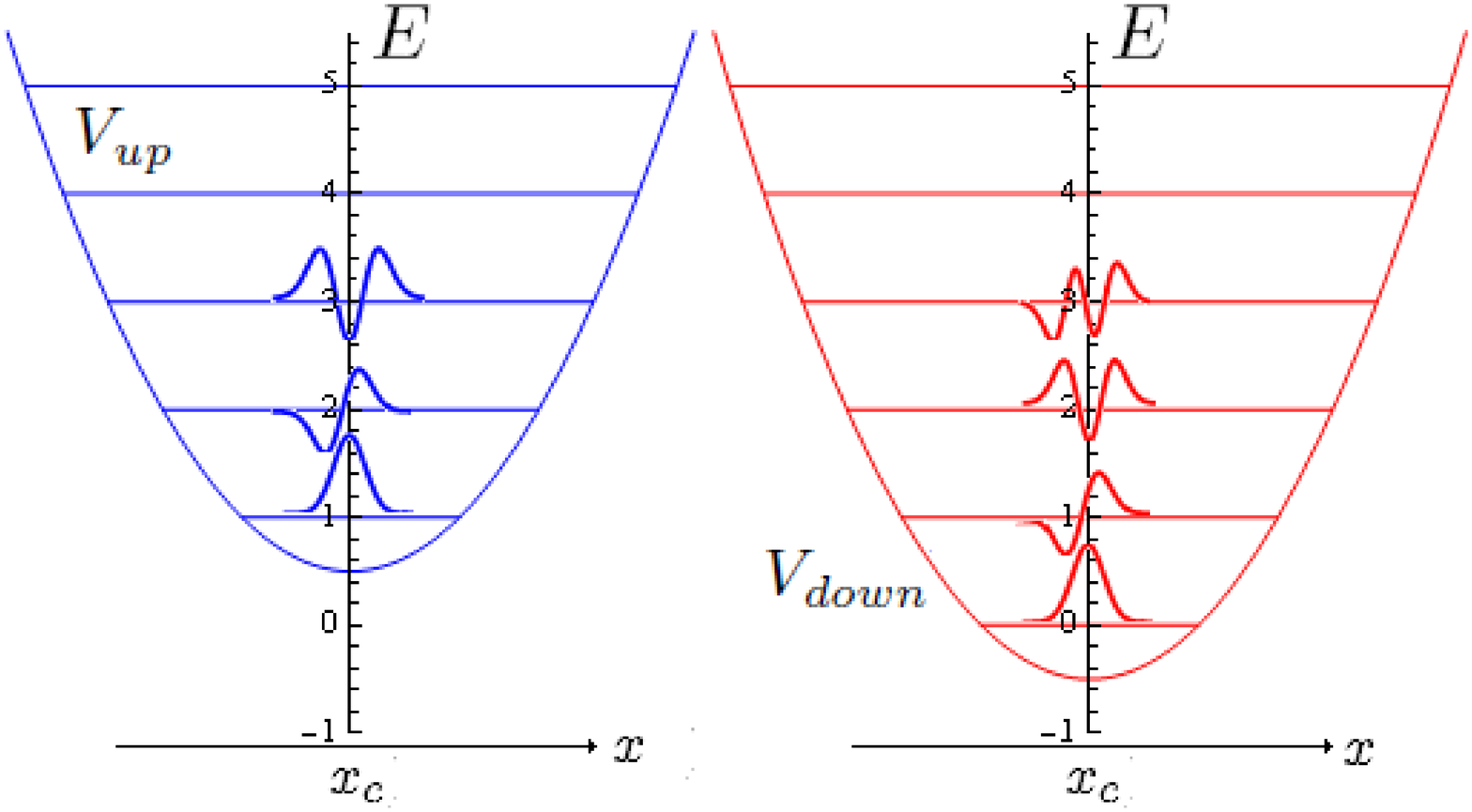}
	\caption{The harmonic potentials $V_{u}$ and $V_{d}$ of the effective Hamiltonian  ${\cal H}_{eff}$. The  eigenfunctions associated with the eigenvalues   $E_n=n$    are the eigenfunctions of the harmonic oscillator, respectively $\Phi_{n-1}$ and $\Phi_n$.}
		\label{fig:zzdp}
\end{figure}

At low energy, the  Hamiltonian (\ref{tb}) can be linearized around the two Dirac points $\K$ and $\Kp$. Expanding $\kk$ as $\kk = \K^{(\prime)} + \q$ in each valley, the tight-binding Hamiltonian is then replaced by a $4 \times 4$ linearized Dirac-like Hamiltonian~:
\begin{equation}
\hat{H}=\gamma a_0\left(
\begin{array}{cccc}
0&i\hat{q}_x-\hat{q}_y& 0 & 0  \\
-i\hat{q}_x-\hat{q}_y& 0 & 0 & 0 \\
0 &0&0 &i\hat{q}_x+\hat{q}_y\\
0 & 0 &-i\hat{q}_x+\hat{q}_y&0
\end{array}
\right)
\label{dirac}
\end{equation}
which describes the two uncoupled valleys in the basis $\left(\varphi_A,\varphi_B,\varphi'_A,\varphi'_B\right)$, where $\gamma=\frac{\sqrt{3}}{2}t$ and $a_0$ is the lattice spacing. The eigenfunctions $\Psi^\pm$ can be expressed as the superposition of the contributions of the two valleys, so that we can write~:
\begin{eqnarray}
\Psi^\pm(\vec{r})=
e^{i \vec{K} \vec{r}}
\left(
\begin{array}{@{}c@{}}
\varphi_A(\rr)\\
\varphi_B(\rr)
\end{array}
\right)
\pm e^{i \vec{K'} \vec{r}}
\left(
\begin{array}{@{}c@{}}
\varphi'_A(\rr)\\
\varphi'_B(\rr)
\end{array}
\right).
\end{eqnarray} 					
In the magnetic field $\vec{B}=-B\vec{e}_z$, and within the Landau gauge $A_y=-Bx$, we perform the Peierls substitution  $\hat{q}_y \rightarrow \hat{q}_y + e\hat{A}_y = \hat{q}_y -eB \hat{x}$.  
Then we introduce the dimensionless variables  $\hat{x}/\ell_B\rightarrow \hat{x}$, and  $x_c= q_y \ell_B$. Since $\hat{q}_x= -i  \partial_{x}$,   the  Hamiltonian in a magnetic field reads~:
\begin{eqnarray}
&&\hat{H}= \frac{\gamma a_0}{\ell_B}\times\\
&&\left(
\begin{array}{@{}c@{}@{}c@{}@{}c@{}@{}c@{}}
0&\partial_x+x-x_c&0&0\\
-\partial_x+x-x_c&0&0 & 0\\
0&0&0&\partial_x-(x-x_c)\\
0&0&-\partial_x-(x-x_c)&0
\end{array}
\right)\notag.
\label{DiracMagn}
\end{eqnarray}
Next, it is useful to work with  the  squared Hamiltonian $\hat{H}^2$ which is diagonal. We  introduce the dimensionless effective Hamiltonian ${\cal H}_{eff}$ as~:
 \begin{equation}
  \hat H^2 = 2 \frac{\gamma^2 a_0^2}{\ell_B^2} {\cal H}_{eff} 
  \label{squareeff}
 \end{equation}
which defines four effective Schr\"odinger equations~:
\be 
{\cal H}_{eff} \Psi = E_n \Psi  
\label{heff1}
 \ee
with
 \be {\cal H}_{eff}=-\frac{\mathbbm{1}}{2} \partial_x^2 \  +
 \left(
 \begin{array}{cccc}
                                                               V_{u}(x) & 0 & 0 & 0 \\
                                                               0 &  V_{d}(x) & 0 & 0 \\
                                                               0 & 0 &  V_{d}(x) & 0 \\
                                                               0 & 0 & 0 &  V_{u}(x) \\
                                                             \end{array}
                                                             \right)
                                                             \label{heff}
                                                              \ee
where the potentials $V_{u}(x)$ and $V_{d}(x)$  are given by~:
\begin{eqnarray}
\label{potup}
V_{u}(x)&\equiv &\frac{1}{2}(x-x_c)^2+\frac{1}{2}\\
V_{d}(x)&\equiv &\frac{1}{2}(x-x_c)^2-\frac{1}{2} 
\label{potdown}
\end{eqnarray}
and are displayed in Fig \ref{fig:zzdp}. Thus, the effective Hamiltonian ${\cal H}_{eff}$ simply describes only two different harmonic oscillators. Because of the non-commutativity of the operators $\hat{x}$ and $\partial_x$, the two harmonic potentials $V_u(x)$ and $V_d(x)$ are shifted by one energy level. From (\ref{heff1}) and (\ref{heff}), one obtains two different Schrödinger equations, one with the potential $V_u(x)$ for $\varphi_A$ and $\varphi'_B$, and the other one with the potential $V_d(x)$ for $\varphi_B$ and $\varphi'_A$. 

The eigenvalues $\ep_n$ of the original Hamiltonian (\ref{DiracMagn}) are obtained from the dimensionless eigenvalues $E_n$ of the effective Hamiltonian ${\cal H}_{eff} $ as~:
  \be 
  \ep_n= \pm  \frac{\gamma a_0}{\ell_B} \sqrt{2E_n}.
   \label{en} 
   \ee
  In the Landau gauge we have used, the wave function is a plane wave along the $y$ direction and now reads~:
 \begin{eqnarray}
 \begin{array}{rl}
\Psi^\pm(\vec{r})=&
\left(
\begin{array}{@{}c@{}}
\Psi^\pm_A(\vec{r})\\
\Psi^\pm_B(\vec{r})
\end{array}
\right)\\
&=e^{i q_yy}
\left[
e^{i \vec{K} \vec{r}}
\left(
\begin{array}{@{}c@{}}
\varphi_A(x)\\
\varphi_B(x)
\end{array}
\right)
\pm e^{i \vec{K'} \vec{r}}
\left(
\begin{array}{@{}c@{}}
\varphi'_A(x)\\
\varphi'_B(x)
\end{array}
\right)
\right].
\label{psi}
\end{array}
\end{eqnarray}
The eigenvalues $E_n=n$ are the Landau levels associated with the  eigenfunctions of the harmonic oscillator $\left\{\Phi_n\right\}$ as follows~:
\begin{eqnarray}
\begin{array}{ll}
\varphi_A=\Phi_{n-1} &\  \varphi'_A=\Phi_n\\
\varphi_B=\Phi_n      &\ \varphi'_B=-\Phi_{n-1}
\end{array}
\label{ohzz}
\end{eqnarray}
This is illustrated in Fig. \ref{fig:zzdp}. The effect of the edges is to modify the potentials (\ref{potup}) and (\ref{potdown}) and consequently  the components $\varphi^{(\prime)}_{A/B}$ of the wave functions and the energy spectrum. These modifications depend on the nature of the edge which yields to specific boundary conditions. In the two following subsections, we derive the effective Schrödinger equations for zigzag and armchair edges.

																										 \subsection{Zigzag edges}
                                                      \label{sec:zz}
\begin{figure}[!ht]
	\centering
		\includegraphics[width=9cm]{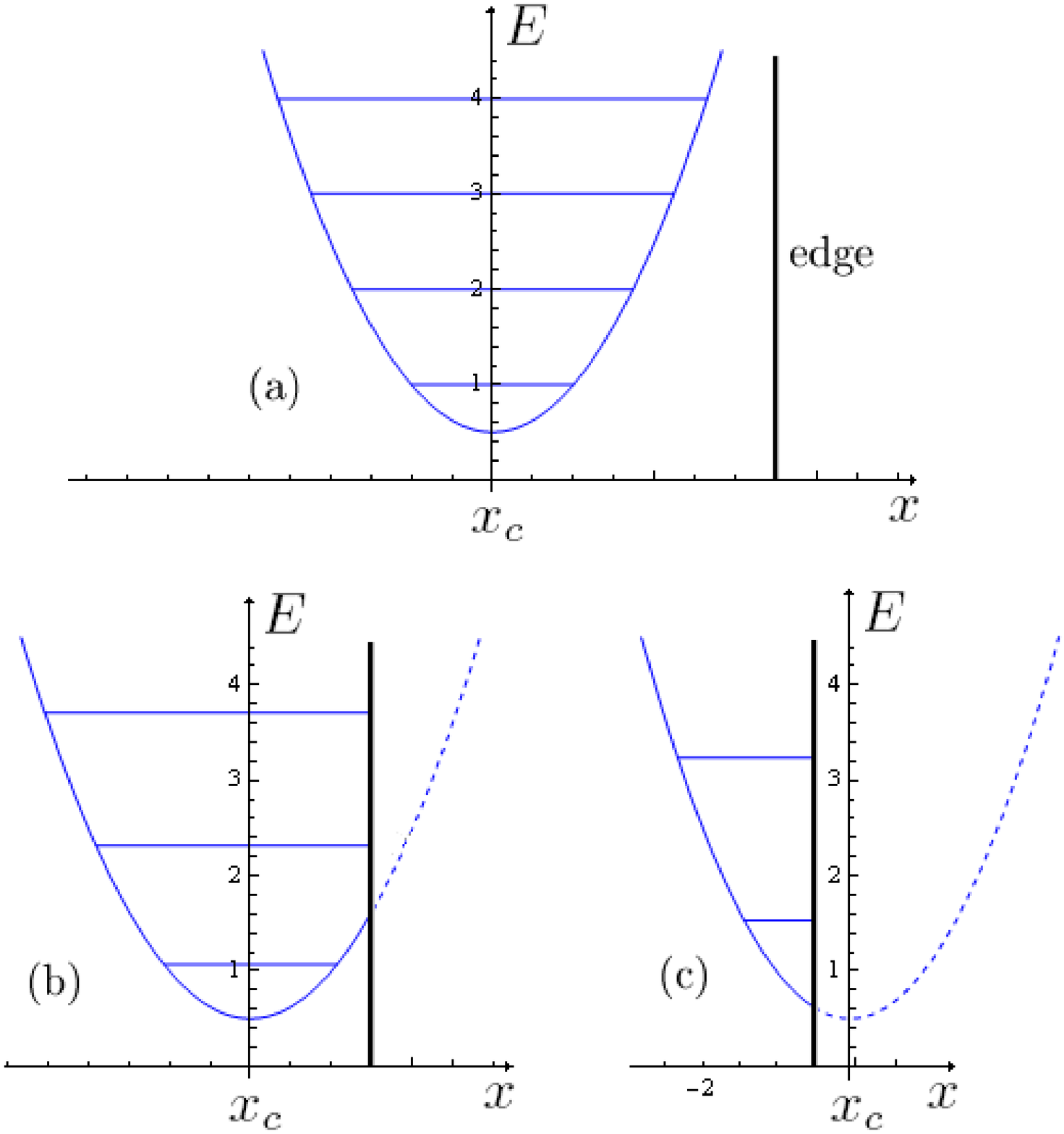}
	\caption{Potential $V_{A}(x)$ that describes the right edge in the $\K$ valley of a zigzag nanoribbon for the ${\cal{H}}_{eff}$ problem. The edge is modelized by an infinite potentia barrier.}
		\label{fig:sozz}
\end{figure}

\begin{figure}[!ht]
	\centering
		\includegraphics[width=9cm]{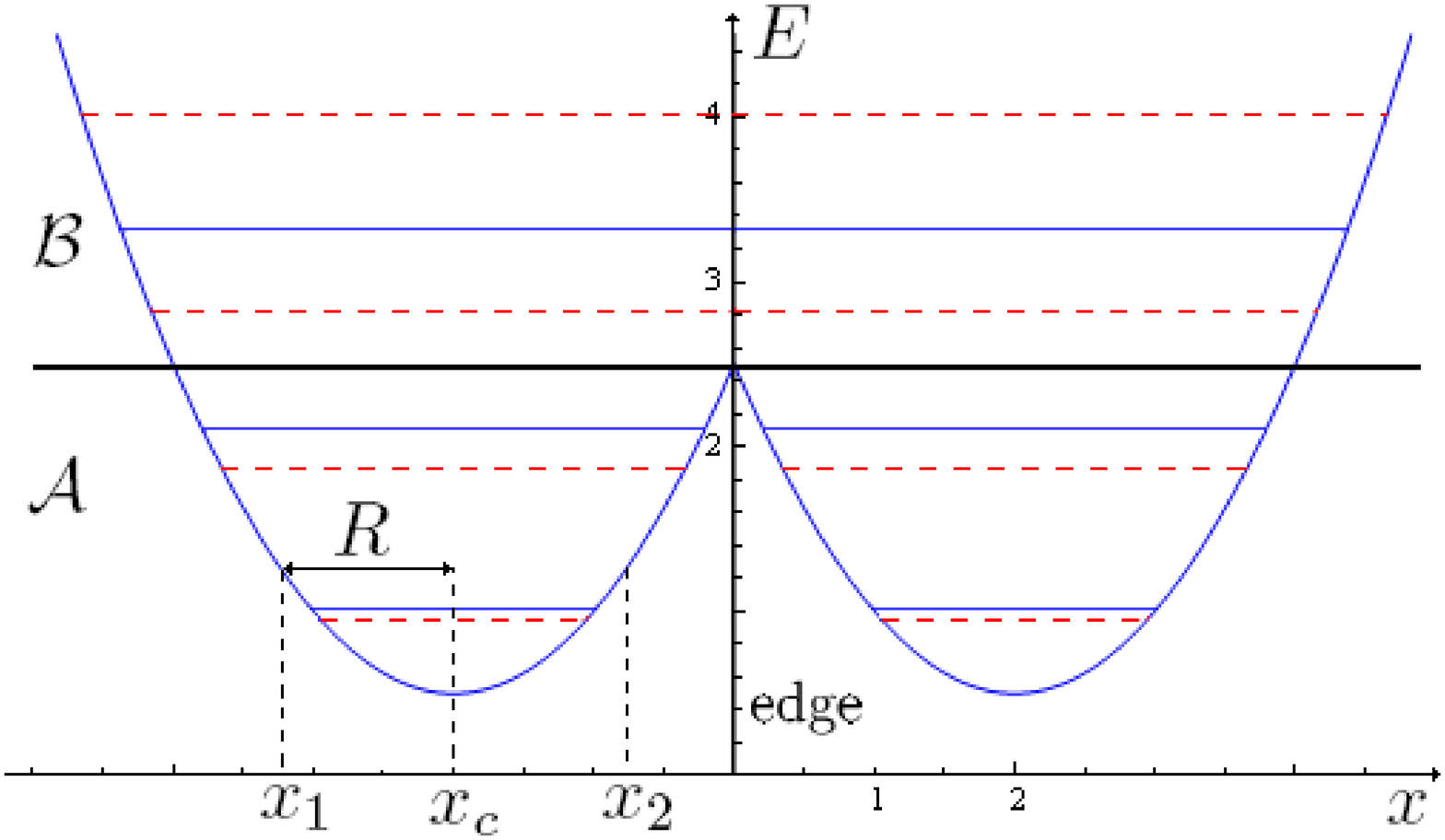}
	\caption{Double symmetric harmonic potential with $x_c=L-2$. We show the eigenenergies obtained within the WKB approximation.  The higher energy levels (blue continuous lines) are the antisymmetric solutions. They are the same than those of  a harmonic well cut by an infinite potential barrier (Fig. \ref{fig:sozz}). For completion we have indicated the energies of the symmetric solutions (dashed red lines) which are nore considered here. The two regions $\cal A$ and $\cal B$ and the cyclotron radius $R$ are defined in  section \ref{sec:sc}.}
		\label{fig:dpsym}
\end{figure}

To treat the zigzag  edges of a graphene ribbon, we first recall that all the atoms on one edge belong to the same sublattice, and therefore, all the atoms on the opposite edge necessarily belong to the other sublattice. As the ribbon is periodic in the $y$ direction, we still write the low energy wave functions $\Psi^\pm$ as in (\ref{psi}) but the components $\varphi^{(\prime)}_{A/B}$ are not the eigenfunctions of the harmonic oscillator anymore. As the left edge is only made of A sites (see Fig. \ref{fig:rubans}), the wave function on  the B sites   vanishes for $x=0$ i.e. $\Psi^\pm_{B}(x=0,y)=0$. The situation is identical on the right edge at $x=L$ for the A sublattice : $\Psi^\pm_{A}(x=L,y)=0$. As a  consequence, we have the boundary conditions~:
\begin{eqnarray}
\begin{split}
\varphi_A(L)=0 & \varphi'_A(L)=0\\
\varphi_B(0)=0 & \varphi'_B(0)=0  .
\end{split}
\label{bczz}
\end{eqnarray}
These boundary conditions do not admix the valleys which can still be described separately, as in the bulk system. Consequently, with the zigzag boundary conditions, we have four independent Schr\"odinger equations, one for each valley and for each sublattice, with four independent potentials $V_A(x)$, $V_B(x)$, $V'_A(x)$ and $V'_B(x)$. Such boundary conditions can be accounted for  by an infinite potential barrier either at $x=0$ or at $x=L$, acccording to the sublattice and the valley. Namely, the potentials $V_B(x)$ and $V'_B(x)$ are  harmonic potentials that have  to be cut in $x=0$ whereas the potentials $V_A(x)$ and $V'_A(x)$ have to be cut in $x=L$. Since the four edge problems are quite similar, we focus on the case of the right edge for the $\K$ valley. Thus, the potential $V_A(x)$ reads~:
\be
V_A(x)=
  \begin{cases}
  \begin{array}{ll}
   V_{u}(x)&\text{for}\quad x<L \\
   \infty &\text{for} \quad x>L
  \end{array}
  \end{cases}
  \label{potzz}
\ee
and is plotted in Fig. \ref{fig:sozz}. Similarly, one can define $V'_A(x)$, $V_B(x)$ and $V'_B(x)$.
As a matter of fact, the problem of a harmonic well with an infinite  potential barrier (Fig. \ref{fig:sozz}) is identical to the problem of a double symmetric harmonic well (Fig. \ref{fig:dpsym}), provided we only keep, in the latter, the eigenenergies associated with the eigenstates that vanish on the edge, that are the antisymmetric states. The interest of considering such a double symmetric well is that, as we will see in section \ref{secarmchair}, both zigzag and armchair edges can be described by double harmonic wells, the difference being an energy shift between the two wells in the armchair case. We illustrate in Fig. \ref{fig:dpsym} the double symmetric potential $V_A(x)$ we finally consider in the effective Schr\"odinger equation for the right zigzag edge and the $\K$ valley, that is~:
\be
&&\left(-\frac{1}{2} \partial_x^2 +V_A(x)\right)\Phi_{AS}(x)=E_n^{AS}(x_c)\Phi_{AS}(x)\\
&&V_A(x)=\frac{1}{2}(|x|+x_c)^2+\frac{1}{2}\ .
\label{potzz}
\ee
The index $AS$ refers to the antisymmetric solutions and we have performed the translations $x-L\rightarrow x$ and $x_c-L\rightarrow x_c$, so that the right edge is now located in $x=0$. The eigenenergies $E_n^{AS}(x_c)$ are calculated analytically within a WKB treatment introduced in section \ref{sec:sc}. The detailed calculations are explained in the appendix.

																									 \subsection{Armchair edges}
																										 \label{secarmchair}																																 
\begin{figure}[h!]
	\centering
		\includegraphics[width=9cm]{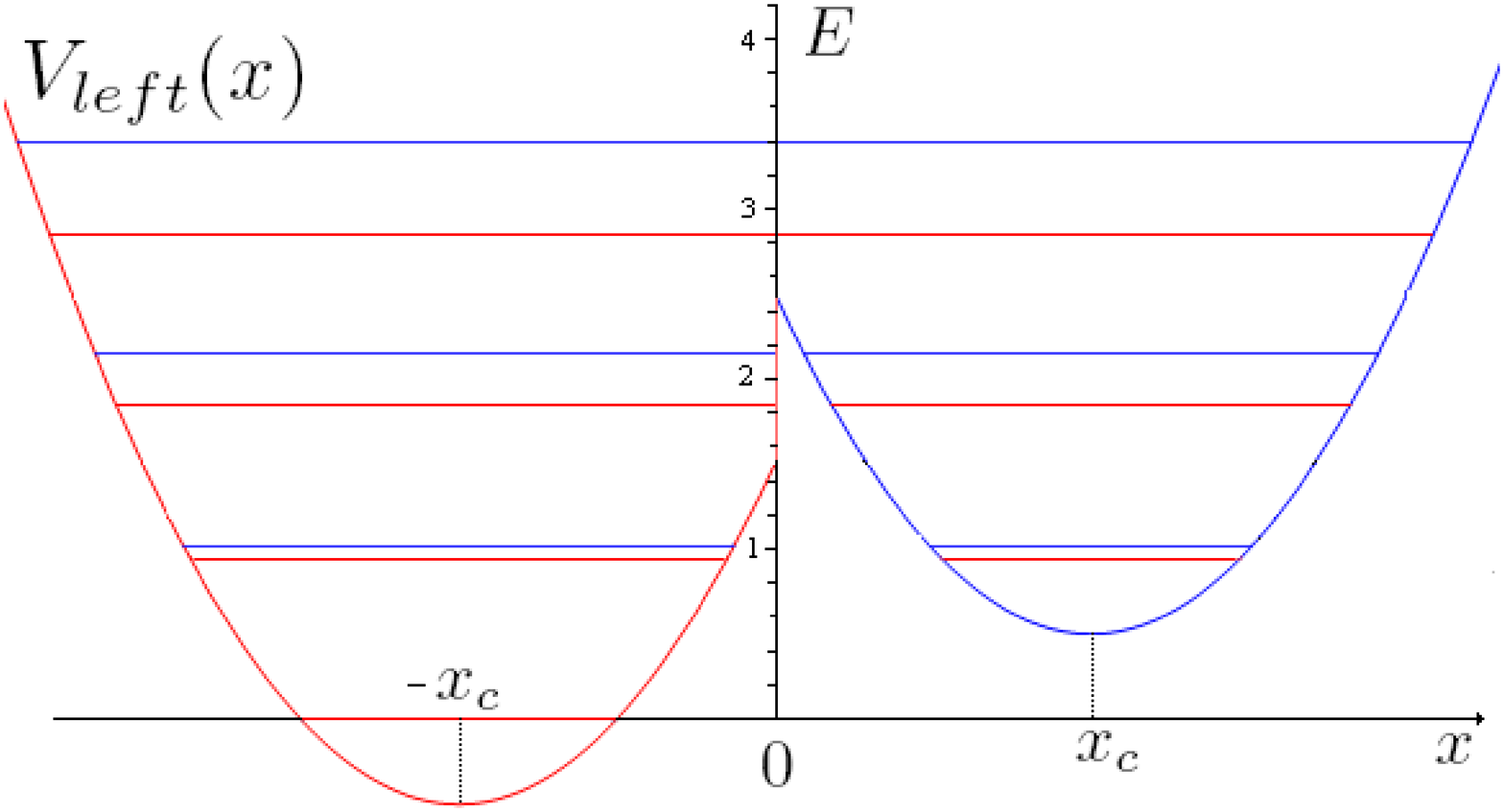}
	\caption{Potential $V_{left}(x)$  for the left edge of an armchair ribbon in a magnetic field. The two wells are centered from a distance $|x_c|$ to the  edge $x=0$. The energy  levels $E_n^+$ in blue and $E_n^-$ in red are shown by horizontal lines in each well and have been calculated semiclassically. The energies are dimensionless. }
		\label{fig:pot}
\end{figure}

We now consider the case of an armchair ribbon as displayed in Fig. \ref{fig:rubans}(b).	 The corresponding   low energy Hamiltonian is obtained from (\ref{dirac}) by the substitution $q_y\rightarrow -q_x$ and $q_x\rightarrow q_y$. By keeping the same gauge $A_y=-Bx $ as above, the Hamiltonian in the presence of a magnetic field now reads~:
\begin{eqnarray}
\label{diracBarm}
&&\hat{H}= i \frac{\gamma a_0}{\ell_B}\times\\
&&\left(\begin{array}{@{}c@{}@{}c@{}@{}c@{}@{}c@{}}
0&-\partial_x-(x-x_c)&0&0\\
-\partial_x+ x-x_c&0&0 & 0\\
0&0&0&\partial_x-(x-x_c)\\
0&0&\partial_x+ x-x_c&0
\end{array}
\right)\notag.
\end{eqnarray}
Of course, this new choice of axes does not affect the eigenvalues of the bulk problem and leaves the effective Hamiltonian ${\cal{H}}_{eff}$ defined in (\ref{squareeff}) unchanged, but the eigenfunctions have now the following form~:
\begin{eqnarray}
\begin{split}
\varphi_A=-i\Phi_{n-1}&\qquad  \varphi'_A=i \Phi_n\\
\varphi_B=\Phi_n      &\qquad \varphi'_B=-\Phi_{n-1}
\end{split}
\label{oh}
\end{eqnarray}
where the $\left\{\Phi_n\right\}$ functions are the eigenfunctions of the harmonic oscillator. The phase factor $i$ originates from the orientation of the ribbon. Following the procedure suggested by Brey and Fertig,\cite{bf} we now construct an effective Hamiltonian which accounts properly for the armchair boundary conditions. Since, for this type of edge,  we have the particular relation $K_y=K^{'}_y$, as argued in section \ref{tbs}, the  wave functions now read~:
\begin{eqnarray}
\Psi^{\pm}(\vec{r})&=&
\left(\begin{array}{@{}c@{}}
\Psi^\pm_A(\vec{r})\\
\Psi^\pm_B(\vec{r})
\end{array}\right)\\
&=&e^{i (K_y+q_y) y}\left[
e^{i K_x x}\left(
\begin{array}{@{}c@{}}
\varphi^\pm_A(x)\\
\varphi^\pm_B(x)
\end{array}
\right)
\pm e^{i K'_x x}
\left(
\begin{array}{@{}c@{}}
\varphi'^\pm_A(x)\\
\varphi'^\pm_B(x)
\end{array}
\right)\right]\notag
\end{eqnarray}
where the components $\varphi^{(\prime)\pm}_{B/A}$ must satisfy specific boundary conditions. A crucial difference with  zigzag ribbons is that  armchair edges are constituted of both A and B sites. As a consequence, {\em both} components  $\Psi_A^\pm$ and $\Psi_B^\pm$ must  vanish on each side of the ribbon:  $\Psi^\pm(x=0,y)=0$ and $\Psi^\pm(x=L,y)=0$. From now on, we choose to explicit these boundary conditions for the left side at $x=0$, what leads to~:
\begin{equation}
\begin{split}
\varphi^\pm_A(0)=\mp \varphi'^\pm_A(0)\\
\varphi^\pm_B(0)=\mp \varphi'^\pm_B(0).
\end{split}
\label{cla}
\end{equation}
Contrary to the zigzag case,  the armchair boundary conditions admix the contributions of the two valleys for a given sublattice. In addition, since the Dirac Hamiltonian is first order, the continuity equations  (\ref{cla}) on the wave function imply continuity of the derivatives. From   (\ref{diracBarm}) and (\ref{cla}) we obtain~:
\begin{equation}
\begin{split}
\partial_x\varphi^\pm_A \Big|_0&=\pm \partial_x\varphi'^\pm_A \Big|_0\\
\partial_x\varphi^\pm_B\Big|_0&=\pm \partial_x\varphi'^\pm_B \Big|_0  \ .
\end{split}
\label{cl2b}
\end{equation}
Next, as suggested by Brey and Fertig,\cite{bf} we build  new functions $\Phi^\pm$ as~:
\begin{eqnarray}
\Phi^{\pm}(x) &\equiv& \varphi^\pm_B(-x)\theta(-x) \mp \varphi'^\pm_B(x)\theta(x)\notag\\
              &=& -i\left[\varphi'^\pm_A(-x)\theta(-x) \mp \varphi^\pm_A(x)\theta(x)\right] 
\label{PhiAB}
\end{eqnarray}
where $\theta$ is the Heaviside function. Thus, the functions $\Phi^\pm$  are solutions of a new effective Schr\"odinger equation with a potential $V_{left}(x)$ which is $V_u(x)$ for $x>0$ and $V_d(x)$ for $x<0$~:
\begin{eqnarray}
\label{searm}
 \left(-\frac{1}{2}\partial^2_x+V_{left}(x)\right)\Phi^{\pm}(x)=E_n^{\pm}(x_c) \Phi^{\pm}(x)\\
V_{left}(x)=\frac{1}{2}\left[\left( |x|- x_c\right)^2+\theta(x)-\theta(-x)\right]
\label{V}
\end{eqnarray}
In order to take into account the armchair boundary conditions (\ref{cla}-\ref{cl2b}), we impose the functions $\Phi^{\pm}_{A/B}$ to satisfy the continuity equations~: 
\be
\begin{array}{l}
\Phi^\pm(0^+)=  \Phi^\pm(0^-) \\
\partial_x \Phi^\pm \Big|_{0^+} =\partial_x \Phi^\pm \Big|_{0^-} .
\end{array}
\label{clPhi}
\ee
The   asymmetric   potential (\ref{V}) is shown in Fig. \ref{fig:pot}. The asymmetry originates from the non-commutativity of $x$ with $\partial_x$ and corresponds to a  shift in energy by one Landau level between the two uncoupled wells $V_{u}$ and $V_{d}$. The asymmetric structure of $V_{left}$ is a consequence of the valley admixing imposed by the boundary conditions.

\begin{figure}[h!]
	\centering
		\includegraphics[width=9cm]{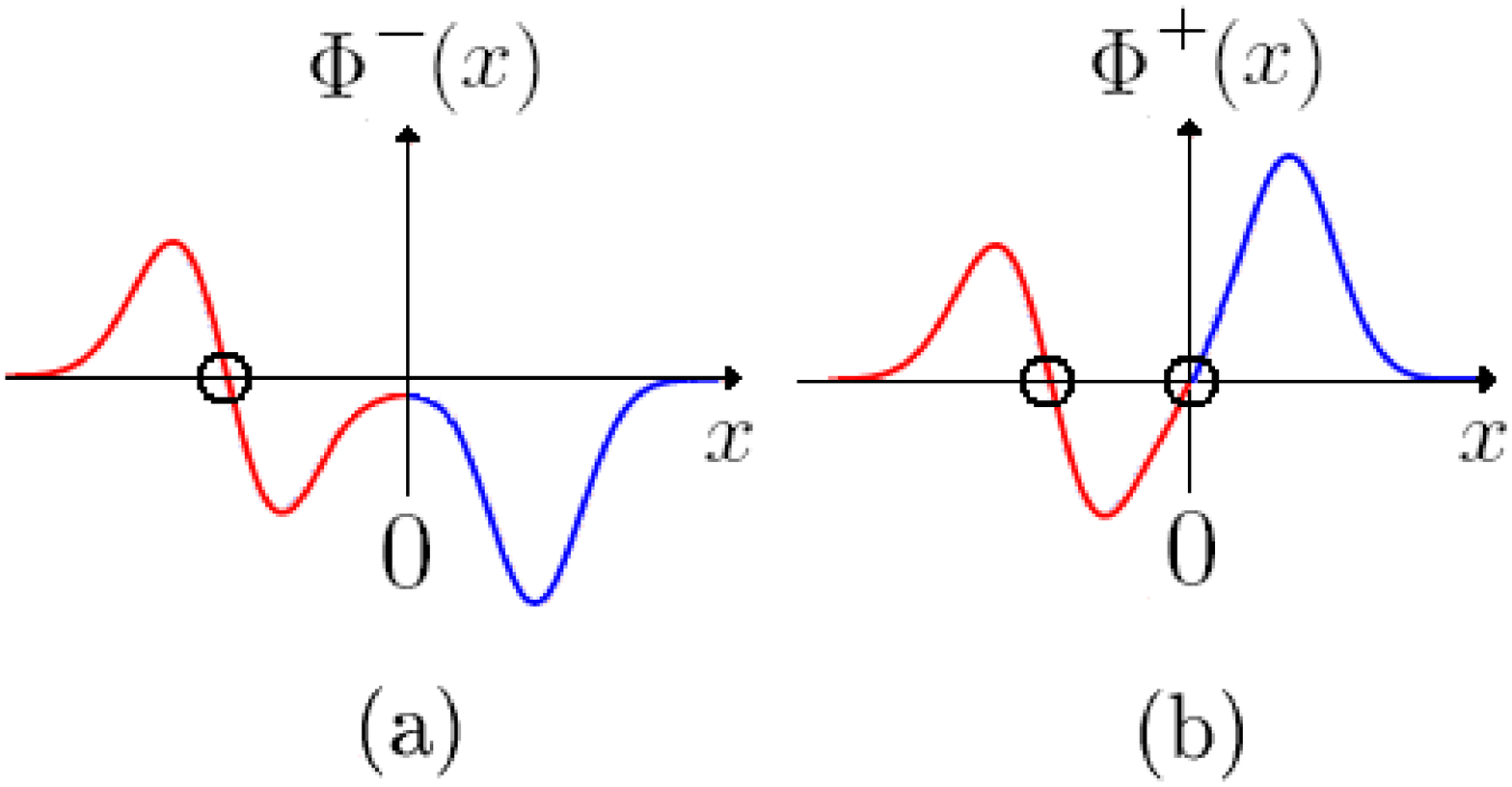}
	\caption{Illustration, for $n=1$,    of the two solutions $\Phi^{-}(x)$ and $\Phi^{+}(x)$ at the left armchair edge. These solutions do not exhibit the same number of nodes (black circles). (a) The continuity equations generate a wave function $\Phi^-$ with $2n-1$ nodes. The fonctions $\varphi_B^-(-x)$ and $-i\varphi_A^{\prime -}(-x)$ are represented on the left (red) whereas $\varphi_B^{\prime -}(x)$ and $-i\varphi_A^{ -}(x)$ are represented on the right (blue). (b) The continuity equations  generate a wave  function $\Phi^+$ with $2n$ nodes. The functions $\varphi_B^+(-x)$ and $-i\varphi_A^{\prime +}(-x)$ are represented on the left (red) whereas $\varphi_B^{\prime +}(x)$ and $-i\varphi_A^{ +}(x)$ are represented on the right (blue).}
		\label{fig:ds}
\end{figure}

We now discuss the qualitative structure of the wave functions. The two combinations  $\Psi^\pm$ are built with four components $\varphi_{A/B}^{(\prime)\pm}$. When the particule is far from the edge, the system is described by two independent potentials $V_u(x)$ and $V_d(x)$ so that the components $\varphi_{A/B}^{(\prime)\pm}$ take their bulk value given by (\ref{oh}). Now, close to the edge, the expressions of $\varphi_{A/B}^{(\prime)\pm}$ are modified. In particular, the boundary conditions (\ref{clPhi}) imply two types of solutions $\Phi^+$ and $\Phi^-$, by matching the components $\varphi_{A/B}^{(\prime)\pm}$ in two different ways. We illustrate this point in Fig. \ref{fig:ds} in the case where $n=1$ by showing a qualitative construction of the components $\varphi^{(\prime)\pm}_{A/B}$. The eigenfunction $\Phi^+$ has $2n$ nodes whereas $\Phi^-$ has $2n-1$ nodes.  As a consequence, the associated eigenvalues $E_n^-(x_c)$ are lower than the eigenenergies $E_n^+(x_c)$, and the valley degeneracy of the energy spectrum is lifted, as seen in Fig. \ref{fig:champarm}. This point is discussed quantitatively in section \ref{sec:results}.  

\begin{figure}[h!]
	\centering
		\includegraphics[width=9cm]{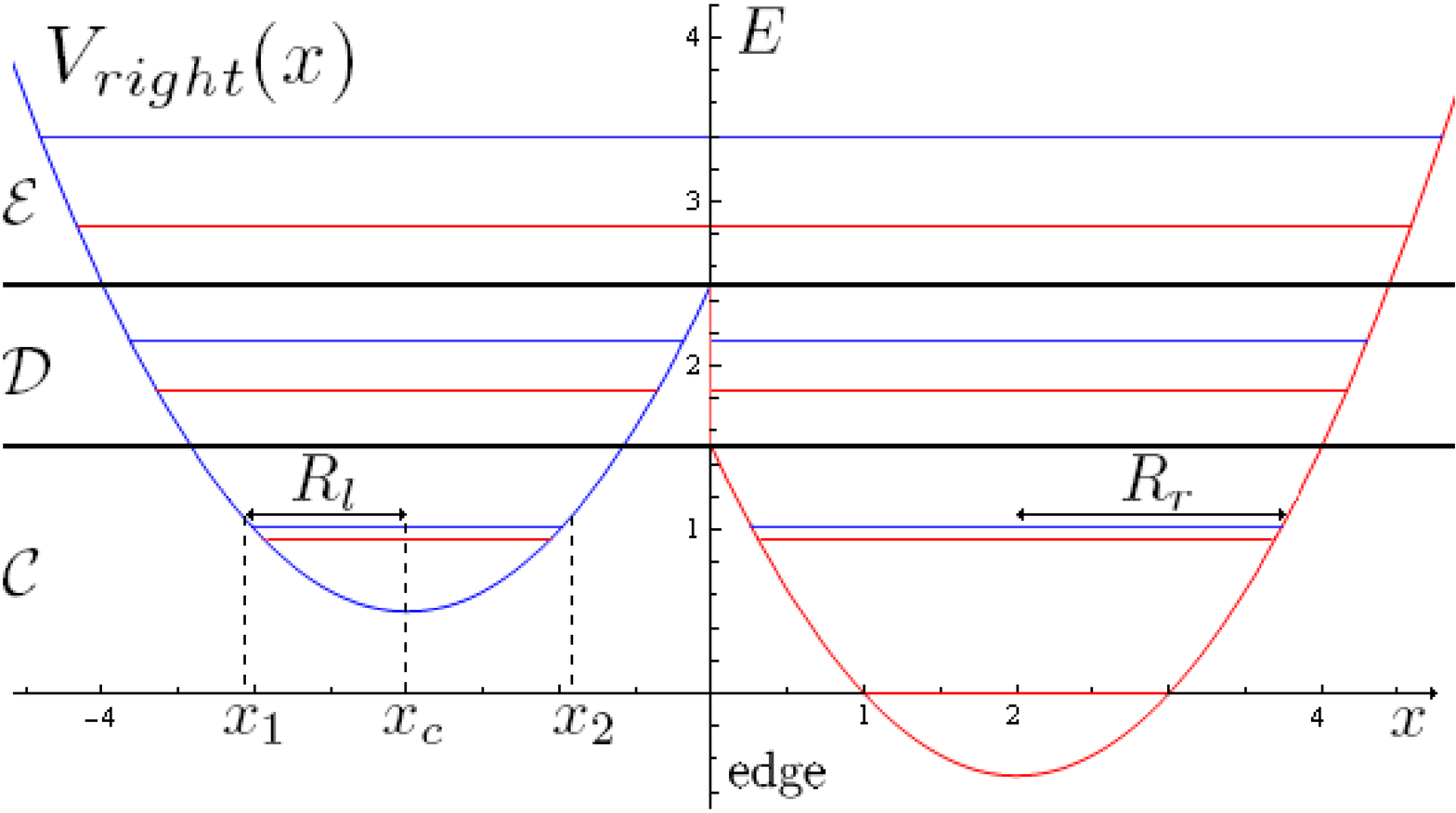}
	\caption{Potential $V_{right}(x)$ of the right armchair edge in a magnetic field with the dimensionless energies $E_n$ calculated semiclassicaly. Here $x_c=-2$. The three regions $\cal C$, $\cal D$ and $\cal E$, such as the cyclotron radii $R_l$ and $R_r$ refer to the section \ref{sec:sc}.}
		\label{fig:armpot}
\end{figure}

By imposing $\Psi(L,y)=0$, the right edge can now be considered easily in the same way. We obtain a similar effective problem where the potential $V_{left}(x)$ has been replaced by~:
\be
V_{right}(x)=\frac{1}{2}\left[\left(\left|x\right|+ x_c\right)^2+\theta(-x)-\theta(x)\right]
\label{potarm}
\ee
where we have performed the translations $x-L\rightarrow x$ and $x_c-L\rightarrow x_c$ for more commodity in the calculations in the next section. Thus,  the potential $V_{right}(x)$ that describes the right edge at $x=L$ is now centered in $x=0$. The potential $V_{right}(x)$ is displayed in Fig. \ref{fig:armpot}.

                            \section{Semiclassical treatment}
                            \label{sec:sc}

The aim of this section is to calculate analytically the energy spectrum $E_n(x_c)$ for both zigzag and armchair boundary conditions, by using a semiclassical formalism. The eigenenergies are solutions of effective Schr\"odinger equations in appropriate potentials: a double symmetric harmonic potential $V_A(x)$ (Fig. \ref{fig:dpsym}) for the zigzag case and a double asymmetric harmonic potential $V_{right}(x)$ (Fig. \ref{fig:armpot}) for the armchair case. We develop here two approaches. In the first one (Sec. \ref{sec:scqa}), we assume that the two wells of the potentials are uncoupled, and quantize the action with the Bohr-Sommerfeld rule.  The second approach (Sec. \ref{sec:wkb}), based on the WKB formalism, properly accounts for the overlap of the wave function between the two wells. 

From now on, we only focus on the right edge of the ribbons, and take $\hbar=1$.


                            \subsection{Semiclassical quantization of the action}
                             \label{sec:scqa}

We first define the classical cyclotron radius $R$. For the zigzag case, $R$ is given  by $E=1/2(R^2+1)$ as illustrated in Fig. \ref{fig:dpsym}. For the armchair case, since the potential $V_{right}(x)$ is asymmetric, we need to define two classical cyclotron radii $R_l$ and $R_r$ (see Fig. \ref{fig:armpot}). Next, we introduce an action $S$ associated with these potentials. Along a closed path, the action is given by~:
\begin{equation}
S(E,x_c)= 2\int_{x_1}^{x_2} dx \ \sqrt{2 (E-V(x))}
\label{action}
\end{equation}
where $x_1$ and $x_2$ are the positions of the turning points, and $V(x)$ is either $V_A(x)$ or $V_{right}(x)$. Such an action is quantized with the Bohr-Sommerfeld rule as~:
\begin{equation}
S(E,x_c) = 2\pi  \left(n+\gamma(E,x_c)\right)
\label{bsq}
\end{equation}
where $n\geq 0$ is an integer and $0<\gamma(E,x_c)<1$ is a function that encodes all the information on the connection procedure at the turning points.  For a  harmonic potential, $\gamma=1/2$. Here, for the double well potentials, $\gamma(x_c,E)$ is not a constant anymore, but depends on the energy and on the distance to the edge because of the overlap of the wave function between the two wells. We assume in this section that the two wells are uncoupled, so that we take $\gamma$ as a constant. 

Now we have to specify  distinct   regions in energy, each of them requiring an  appropriate semiclassical treatment. These regions are delimited by horizontal thick lines in Figs. \ref{fig:dpsym} and \ref{fig:armpot} and involve different expressions of the action $S(E,x_c)$ given by (\ref{action}).

																										
																										 \subsubsection{Zigzag ribbons}
																										 
We consider the potential $V_A(x)$ that describes the right zigzag edge for the $\K$ valley (Fig. \ref{fig:dpsym}). We recall that the right edge is located at $x=0$. For a given energy $E$, we have to discuss two distinct regions~: $|x_c|>R$ (region $\cal A$) and $|x_c|<R$ (region $\cal B$), where $R=\sqrt{2E-1}$ is the cyclotron radius. 																								 
\begin{itemize}
	\item Region $\cal A$: $|x_c|>R$. The turning points are defined by $x_1=x_c-R$ and $x_2=x_c+R$ so that the action (\ref{action}) in the left well is simply~:
\begin{equation}
S_{\cal A}(R)=\pi R^2
\label{action1}
\end{equation}
A simple calculation of the energies by quantizing $S_{\cal A}$ with the constant value $\gamma=1/2$ in (\ref{bsq}) leads to the Landau levels $E_n=n+1$ with $n\geq 0$, and therefore to the  spectrum $\ep_n \propto \sqrt{B(n+1)}$ in  this region. The  $n+1$ term originates from the energy shift of the potential $V_A(x)$. The rest of the spectrum near the right edge is given by the contribution of the other valley. Therefore, by treating the same way the potential $V'_A(x)$, we find the spectrum $E_n=n$ with $n\geq 0$. Finally, we obtain a set of degenerated energies $E_n=n$ with $n\geq 1$, and a non degenerated level $E_0=0$ what gives the  expected valley degenerated graphene energy levels  $\ep_n \propto \pm\sqrt{Bn}$ for $n\geq 0$.

\item Region $\cal B$: $|x_c|<R$. The left turning point $x_1$ is unchanged but the right turning point is replaced by $x_2=0$ so that the action reads~:
\begin{eqnarray}
S_{\cal B}(R,x_c)= R^2\left(\theta-\frac{1}{2}\sin(2\theta)\right)
\label{action2}
\end{eqnarray}
where we have introduced the parameter $\theta \equiv \arccos{\frac{x_c}{R}}$. The spectrum depends now on the distance $x_c$ to the edge. The total action in the double well $S_t=2S_{\cal B}$ is quantized as $S_t=2\pi(p+1/2)$. We recall that we have to keep only the antisymmetric solutions which have an odd number of nodes, what implies $p=2n+1$. Therefore, such a quantization leads to~:
\be
S_{\cal B}(R,x_c)=2\pi\left(n+\frac{3}{4}\right)\ .
\label{scq2}
\ee   
We recover the result $\gamma=3/4$ for a harmonic potential cut by an infinite barrier potential.\cite{montavi} By identifying the two expressions (\ref{action2}) and (\ref{scq2}), we finally obtain a set of self-consistent equations labeled by the integer $n$ for $E_n(x_c)$, from which we can extract the energy spectrum.
\end{itemize}

The spectrum $E_n(x_c)$ obtained within this approach for the right zigzag edge in the $\K$ valley  is plotted  in Fig. \ref{fig:zzhalperin}. In order to correctly describe the region where $|x_c|\approx R$, we  use in  section \ref{sec:wkb} a more sophisticated approach based on the WKB formalism. 


                                                    \subsubsection{Armchair ribbons}
Because of the asymmetry of the potential $V_{right}(x)$, we have now to distinguish three regions ($\cal C$, $\cal D$, $\cal E$ in Fig. \ref{fig:armpot}). In each one, we consider two actions, $S_l$ for the left well and $S_r$ for the right well. In addition, the levels in the left or in the right well will be indexed by different integers $n_l , n_r \geq 0$. The energy is given by $E=(R_l^2+1)/2=(R_r^2-1)/2$ where $R_l$ (resp. $R_r$) is the cyclotron radius for the left (resp. right) well. 

\begin{itemize}
	\item Region $\cal C$:  $R_l \leq R_r \leq \left|x_c\right| $.  In each well, the action still has the form given by (\ref{action1}),  so that we write~: 
\begin{equation}
\begin{array}{rl}
\text{left well}: & S_l = S_{\cal A}(R_l)\\[3mm]
\text{right well}: & S_r = S_{\cal A}(R_r)\ .
\end{array}
\end{equation}
By using the Bohr-Sommerfeld rule (\ref{bsq})  with $\gamma=1/2$ for both $S_l$ and $S_r$, we find the energies $E_{n_l}=n_l+1$ into the left well and $E_{n_r}=n_r$ into the right well, and then $n \equiv n_r = n_l+1$. This leads to the degenerated Landau levels $E_n=n$ with $n\geq 1$ and a non degenerated level $E_0=0$.

	\item  Region $\cal D$: $R_l \leq \left|x_c\right| \leq R_r$. Now, because of the step in the potential, the action $S_r$ has the form given by (\ref{action2}), whereas $S_l$ is unchanged. The two actions read
	\begin{equation}
\begin{array}{rl}
\text{left well}: & S_l = S_{\cal A}(R_l)\\[3mm]
\text{right well}: & S_r = S_{\cal B}(R_r,x_c)\ .
\end{array}
\end{equation}
The action $S_l$ is still quantized as previously, what simply gives the spectrum $E_n=n$ with $n\geq 1$.  As the wave function inside the right well does not totally vanish in $x=0$, the situation is different than in the case of an infinite potential barrier. This suggests a choice closer to $\gamma=1/2$ rather than $\gamma=3/4$ for the quantization of the action $S_r$. By doing so, we obtain an implicit equation in $E_n$ and $x_c$ from which we can extract the spectrum $E_n(x_c)$. Of course, we understand that the region $\cal D$ is necessarly badly described by such an approach and specifically requires a more sophisticated method since $\gamma$ cannot be a constant anymore.     

	\item  Region $\cal E$: $\left|x_c\right|\leq R_l \leq R_r$. In this region,  we need to consider the total action that reads~:
\begin{equation}
S_{\cal E}=S_{\cal B}(R_l,x_c)+ S_{\cal B}(R_r,x_c)
\end{equation}
where the function $S_B$ is given in (\ref{action2}). This expression is still valid when $x_c$ crosses the edge. Since the two turning points both touch a harmonic potential, this action is simply quantized by the usual Bohr-Sommerfeld rule with $\gamma=1/2$ what also leads to an implicit equation in $E_{n'}$ and $x_c$, where $n'$ is a different integer  than $n$, since there are twice more solutions in this region than in the regions $\cal A$ or $\cal C$.
\end{itemize}

The armchair spectrum obtained within this approach is shown in Fig. \ref{fig:spectrearm}. It perfectly matches with numerical solutions of the Schr\"odinger equation (\ref{searm}) except around the intermediate region $\cal D$.  In the following, we describe quantitatively all the regions by keeping $\gamma=\gamma(E,x_c)$ and calculating the spectra of zigzag and armchair ribbons by accounting for the coupling between the two wells. We give now the general picture of the method which is detailed in the appendix.


                            												\subsection{The WKB approach}
                                                     \label{sec:wkb}
\begin{figure}[!ht]
	\centering
		\includegraphics[width=9cm]{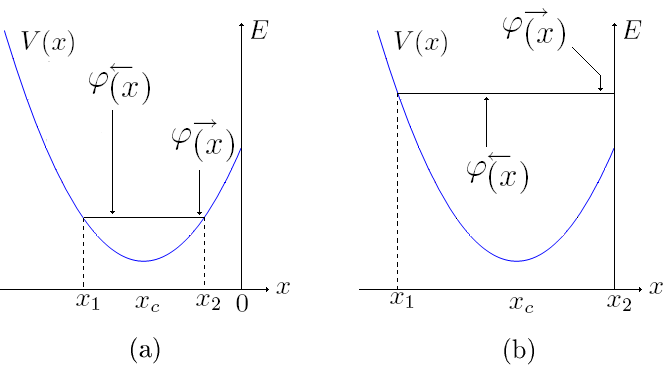}
	\caption{Illustration of the two typical different situations we have to distinguish for the semiclassical calculation. (a) The two turning points $x_1$ and $x_2$ are on the parabola. (b) The turning point $x_2=0$ is not on the parabola. These two situations involve different expressions of the action $S(E,x_c)=S_l$ in the left well. Same considerations have to be made for the right well.}
		\label{fig:wkb}
\end{figure}

In this section, we present a method based on the WKB formalism to account for the overlap of the wave function between the two wells. This approach is detailed in the appendix. Technically, it brings us to calculate the  function $\gamma(E,x_c)$. For this purpose, we express the wave function inside the left well within the WKB approximation as~:
\begin{eqnarray}
\varphi^{\leftarrow}(x)&=& \frac{C}{\sqrt{k(x)}}\sin\left({\cal S}(x_1,x)+\frac{\pi}{4}\right)
\label{wkbt}
\end{eqnarray}
with $C$ a constant and where
\be
{\cal S}(x_1,x)=\int_{x_1}^xdx'\sqrt{E-V(x')}
\ee 
is the partial action between the turning point $x_1$ and an arbitrary position $x$ inside the left well. Close to the right turning point $x_2$, this approximation  breaks down. In order to find a valid approximation of the wave function in this region, we linearise the harmonic potential so that the eigenfunctions of the Schr\"odinger equation are a combination of Airy functions, that is~:
\begin{equation}
\varphi^{\rightarrow}(x) = \alpha_l \mbox{Ai}(f_l(x,E,x_c))+\beta_l \mbox{Bi}(f_l(x,E,x_c))
\label{airy}
\end{equation}
where $\alpha_l$ and $\beta_l$ are constants and $f_l(x,E,x_c)$ is a function that depends on the region $\cal A$, $\cal B$, $\cal C$, $\cal D$ or $\cal E$, so that each region must be treated separately. By imposing the equality between $\varphi^{\leftarrow}(x)$ and the asymptotic expansion of $\varphi^{\rightarrow}(x)$ inside the well, we obtain a relation between the constants and the action $S_l$ which is given by the relation (\ref{action}). We need to distinguish  the cases  $x_2\neq 0$ and $x_2=0$ (Fig. \ref{fig:wkb}), for which the matching condition gives~:
	\begin{equation}
\begin{split}
x_2\neq0:\qquad & \tan{S_l/2}=\alpha_l/\beta_l\\
x_2= 0:\qquad & \tan (S_l/2+\delta_l)=\alpha_l/\beta_l
\end{split}
\label{matchl}
\end{equation}
where $\delta_l=\delta_l(E,x_c)$ is a known function originating from the step between the energy $E$ at the turning point $x_2=0$ and the potential $V(0)$. The same procedure must be performed for the right well where we obtain similar relations between the ratio $\alpha_r/\beta_r$ and the action $S_r$ calculated between the two turning points in the right well~:
 
 	\begin{equation}
\begin{split}
x_3\neq0:\qquad & \tan{S_r/2}=\alpha_r/\beta_r\\
x_3= 0:\qquad & \tan (S_r/2+\delta_r)=\alpha_r/\beta_r
\end{split}
\label{matchr}
\end{equation}
where $x_3$ is the left turning point in the right well.

To determine the ratio $\alpha_l/\beta_l$, we impose the continuity of the wave function and its derivative between the two wells. By doing so, we obtain one equation with the two unknown quantites $\alpha_l/\beta_l$ and $\alpha_r/\beta_r$. Next, by injecting $S_r=S_t-S_l$ into (\ref{matchr}),  where the total action $S_t$ is calculated explicitly as a function of $E$ and $x_c$, we obtain   a second order polynomial in $\alpha_l/\beta_l$ what yields two solutions. Because of the relation (\ref{matchl}), the action  is quantized as $S_{l}=2\pi (n+\gamma_{l})$ which is the Bohr-Sommerfeld quantization. Thus, the two solutions of the  polynomial yield the two solutions $\gamma_l^{\pm}(E,x_c)$.

Next, by using the formula (\ref{action}), we calculate explicitly the expression of the action $S_l$ as a function of   $E$ and $x_c$, which we identify with the action given by the Bohr-Sommerfeld rule previously found. As we know the functions $\gamma_l^{\pm}(E,x_c)$, we finally obtain two analytical self-consistent  equations in $E$ and $x_c$ for a given $n$, one with $\gamma_l^{+}$, the other one with $\gamma_l^-$. The spectrum $E_n(x_c)$ can then be extracted for a given $n$. The whole  procedure must be performed for each region $\cal A$ to $\cal E$ where the action has different expressions as discussed in the previous section. All this study is  detailed in the appendix for both the symmetric and the asymmetric potentials.


                           											 \subsection{Quantitative analytical results}
                           											 \label{sec:results}

We give here the spectra obtained for the effective potentials $V_A(x)$ (zigzag case (\ref{potzz}))  and $V_{right}(x)$ (armchair case (\ref{potarm})),  within the two semiclassical approaches introduced above.
														 											\subsubsection{Symmetric potential (zigzag)}
\begin{figure}[!ht]
	\centering
		\includegraphics[width=9cm]{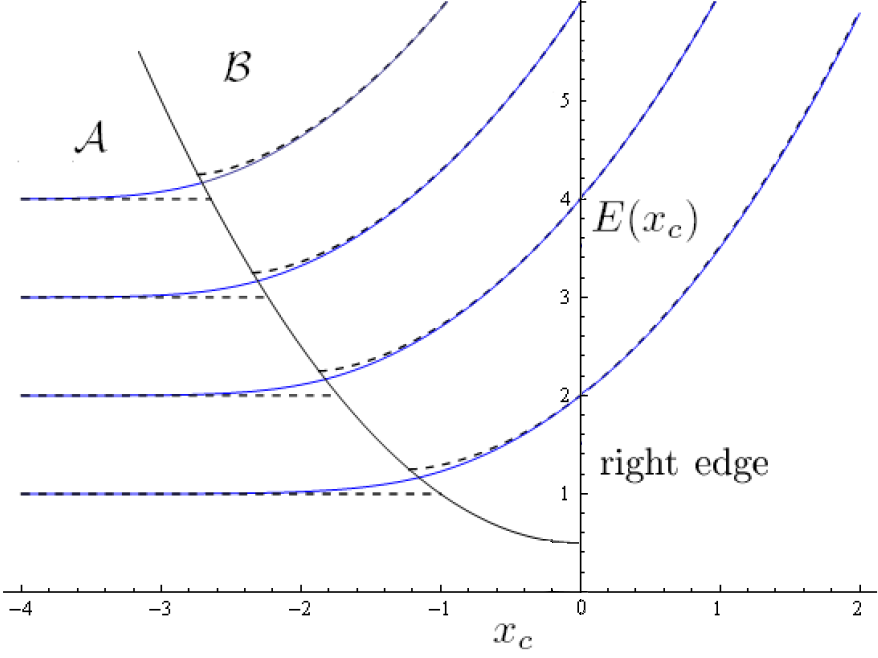}
	\caption{Dimensionless eigenenergies $E_n(x_c)$ of the double symmetric harmonic oscillator potential $V_A(x)$ calculated analytically (continuous line) within the WKB method and (dashed line) within the Bohr-Sommerfeld quantization rule with $\gamma=\frac{1}{2}$ in region $\cal A$ and $\gamma=\frac{3}{4}$ in region $\cal B$. The solutions we keep correspond to the antisymmetric solutions of the double symmetric well.}
		\label{fig:zzhalperin}
\end{figure}

\begin{figure}[!ht]
	\centering
		\includegraphics[width=9cm]{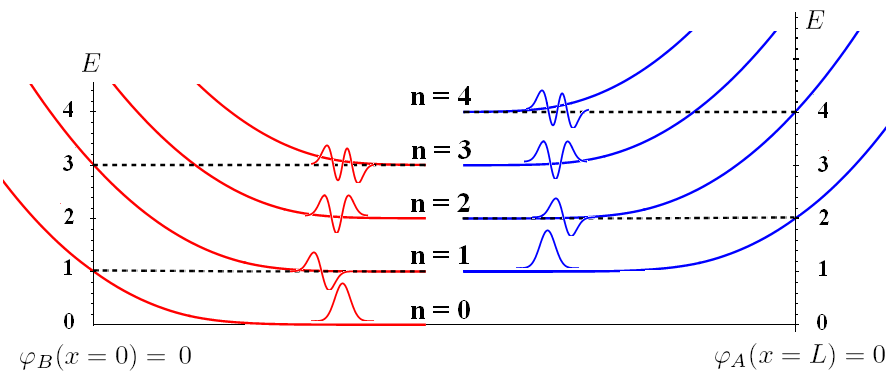}
	\caption{Dimensionless zigzag edge states $E_n(x_c)$ of the ${\cal H}_{eff}$ problem in the $\K$ valley. We have represented the bulk wave function of each sublattice (left (red) for B sublattice and right (blue) for A sublattice). The results have been obtained within the WKB approximation.}
		\label{fig:halperinzz}
\end{figure}

The energy levels $E_n(x_c)$ of the double symmetric harmonic potential $V_A(x)$ describing the contribution of the $\K$ valley at the right zigzag edge (see Fig. \ref{fig:dpsym}) are shown in Fig. \ref{fig:zzhalperin}. The two regions $\left|x_c\right|>R$ and $\left|x_c\right|<R$ are separeted  by a parabola of equation $E=(x_c^2+1)/2$.  In the spectrum of Fig. \ref{fig:zzhalperin}, we have only kept the higher energy solution of the double well problem, since they correspond to the eigenenergies of the antisymmetric eigenfunctions.  The dashed lines correspond to the approximation where $\gamma=cst$, what leads to an unphysical discontinuity around $\left|x_c\right|=R$. The continuous lines represent the energy levels obtained within the WKB approximation. The solutions found with this method in regions $\cal A$ and $\cal B$ perfectly match at $\left|x_c\right|=R$.  
Next, we show in Fig. \ref{fig:halperinzz} the structure of the edge states spectrum in the $\K$ valley for both the left and right edges. The right (blue) part of the spectrum, is the energy spectrum on the right edge calculated within the WKB method and already displayed in Fig. \ref{fig:halperinzz}, whereas the left (red) part represents the edge states on the left side of the ribbon in the same $\K$ valley. The effective problem for the left edge is obtained in a similar way than for the right edge (see section \ref{sec:zz}) where we have considered the potential $V_d(x)$ and the boundary condition $\varphi_B(x=0)=0$.  Note that we recover the particular relation  $\epsilon_n^{edge}=\epsilon_{2n+1}^{bulk}$ with $n\geq 0$  on the left edge  and $\epsilon_n^{edge}=\epsilon_{2n}^{bulk}$  with $n>0$ for the right edge. This peculiar property of the edge states is due to the fact that we have to keep only the antisymmetric eigenfunctions of the double well problem. In particular, when $x_c=0$, the double well is simply a single harmonic well, so as the bulk potenials. In addition, the shift of this distribution between the two edges is due to the shift in energy between the potentials $V_A(x)$ and $V_B(x)$ that respectively originate from the potentials $V_u(x)$ and $V_d(x)$ modified by the zigzag boundary conditions. We have emphasized this point by dashed lines in Fig. \ref{fig:halperinzz}. This explains the  remarkable structure of the zigzag edge states discussed in section \ref{pmf}. Finally, the mirror symmetry between the valleys $\K$ and $\Kp$ emphasized in Fig. \ref{fig:zzx} is now clear, since the edge potentials into the $\Kp$ valley, that are
$V'_A(x)$ for the right edge and $V'_B(x)$ for the left edge, now originate from the bulk potentials $V_d(x)$ and $V_u(x)$ respectively with the zigzag boundary conditions.


																											\subsubsection{Asymmetric potential (armchair)}

\begin{figure}[!ht]	\centering
		\includegraphics[width=9cm]{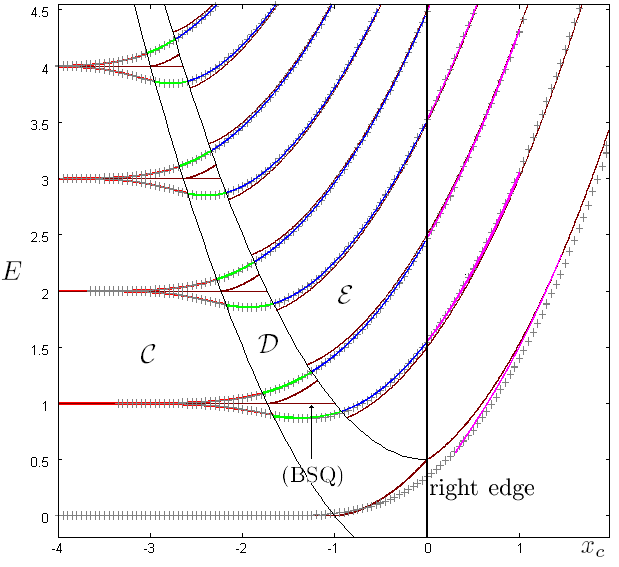}
	\caption{Dimensionless energies $E_n(x_c)$ of the armchair edge states. The results obtained with three different methods are shown. The discontinuous brown lines are the levels obtained by taking $\gamma=cst$ in the Bohr-Sommerfeld quantization (BSQ). The colored continuous lines have been obtained within the WKB approach. The different colors  in regions $\cal C$, $\cal D $ and $\cal E$ are associated to  the different semiclassical behaviours of the particule in Fig. \ref{fig:armso}, as discussed in the next section. The grey crosses represent the numerical results obtained within the tight-binding method. We show that the results obtained within the WKB approximation perfectly fit the numerical calculations, except for $n=0$.}
		\label{fig:spectrearm}
\end{figure}

For the armchair case, we have to keep all the solutions of the double asymmetric potentials $V_{left}(x)$ and $V_{right}(x)$. As discussed in section \ref{secarmchair} and illustrated in Fig. \ref{fig:ds}, the wave functions of the effective Schr\"odinger equation with armchair boundary conditions (\ref{searm}) have, for a given $n$, either  $2n$ nodes or  $2n-1$ nodes, what leads to pairs of eigenenergies $E_n^\pm(x_c)$. The semiclassical  energies $E_n(x_c)$  are displayed in Fig. \ref{fig:spectrearm}. The three regions $\cal C$, $\cal D$ and $\cal E$ are separeted by the parabola parametrized by $\left|x_c\right|=R_l$ and $\left|x_c\right|=R_r$, that are $E=\frac{1}{2}(x_c^2+1)$ and $E=\frac{1}{2}(x_c^2-1)$. Contrary to the simple semiclassical calculation with $\gamma=cst$ (discontinued brown lines), the WKB method gives a perfect matching between the three regions and is in very good agreement with  numerical calculations (grey crosses).


																	   	\section{Quantized skipping orbits}
                                       \label{secarea}
In this section, we propose a simple interpretation of the results in terms of a semiclassical picture for skipping orbits.
Consider the action $S$ associated with the effective Schr\"odinger equations with the potentials obtained for the zigzag case. These four  equations (one per edge and per valley) describe the motion of a free massive particle in a magnetic field (all units being set to one) in the presence of an infinite potential barrier  or the image of the harmonic potential, as illustrated in Figs. \ref{fig:sozz} and \ref{fig:dpsym} for the right edge in the $\K$ valley. From the classical equations of motion, the action $S= \oint \vec{p}  \cdot d\vec{r}$ along a closed trajectory for a given energy is simply related to the area ${\cal A}$ enclosed by the corresponding periodic orbit~:
\be
{\cal S}(E,x_c) = \oint \vec{p}  \cdot d\vec{r} = {e B \over 2 } \oint \vec{r} \times d\vec{r} = e B {\cal A} 
\ee
so that the Bohr-Sommerfeld quantization rule (\ref{bsq}) implies the quantization of the area
\be
 {\cal A}(E,x_c)= 2 \pi (n+ \gamma) \ell_B^2  
   \ee
in our units where $\hbar=1$. In the bulk of the system, that is when the distance $|x_c|$ to the edge is larger than the cyclotron radius $R$, we have simply  
\be {\cal A}(E)= {\cal S}(E)= 2 \pi (n+ \gamma) \ee
and the distances are measured in units of $\ell_B$. This quantization implies the quantization of the cyclotron radius $R$. Taking $\gamma=1/2$, ${\cal A}= \pi R^2= 2 \pi (n+1/2)$, so that $R_n^2= 2 n +1$,   we deduce the energy levels, $E_n=R_n^2/2 \pm 1/2= n$, and obtain the bulk spectrum of Landau levels $E_n=n$.

\begin{figure}[!ht]
	\centering
		\includegraphics[width=9cm]{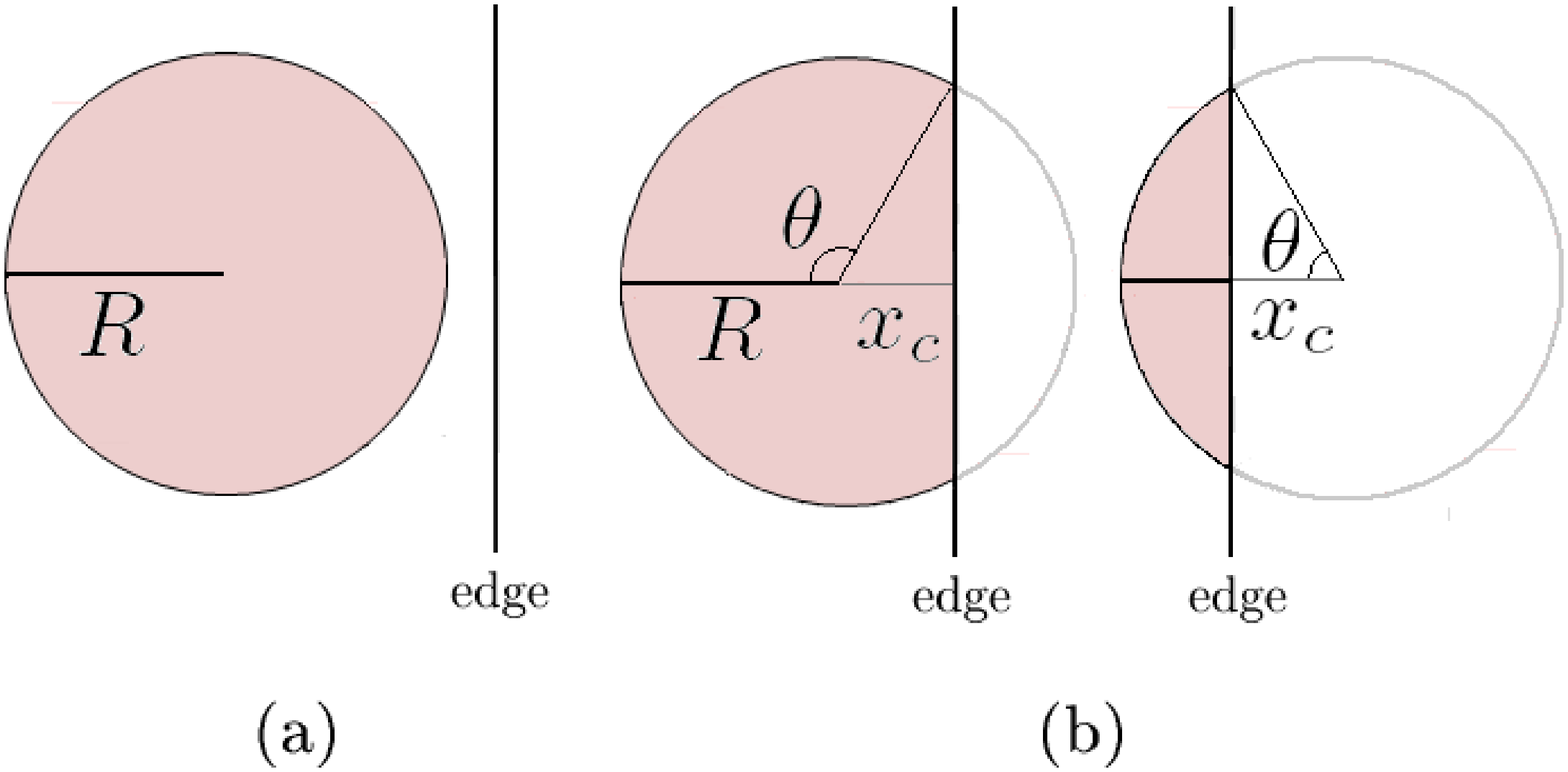}
	\caption{Classical cyclotron orbits  for  the "effective" particle associated to the effective Hamiltonian ${\cal H}_{eff}$ close to a zigzag edge. (a) The closed cyclotron orbit encloses an area ${\cal A}$ quantized by the Bohr-Sommerfeld rule \ref{skipquant}, and the resulting Landau levels spectrum is  displayed in Fig. \ref{fig:zzhalperin} (region $\cal A$). (b) The particle is skipping along the  edge and the edge states energy  spectrum (region $\cal B$ in Fig. \ref{fig:zzhalperin}) is obtained from the quantization rules (\ref{skipquant}).}
		\label{fig:zzso}
\end{figure}

When $|x_c|$ becomes smaller than $R$, the classical orbits skip along the wall represented by the infinite potential well, and they form open orbits. As already noticed\cite{vanhouten}, the area to be quantized is the area delimited by the skipping orbit and the wall. In this case, the action calculated in (\ref{action2}) has a very simple interpretation in terms   of skipping orbits since the parameter $\theta= \arccos{\frac{x_c}{R}}$ has a   clear geometrical meaning shown in Fig. \ref{fig:zzso}. The action is nothing but the area enclosed by the skipping orbit~:
\begin{eqnarray}
{\cal A}(E,x_c)=R^2\left[\theta-\frac{1}{2}\sin(2\theta)\right]
\label{aire2}
\end{eqnarray}
We can now interpret semiclassically the energy spectra obtained in the above subsections.

\begin{figure}[!ht]
	\centering
		\includegraphics[width=9cm]{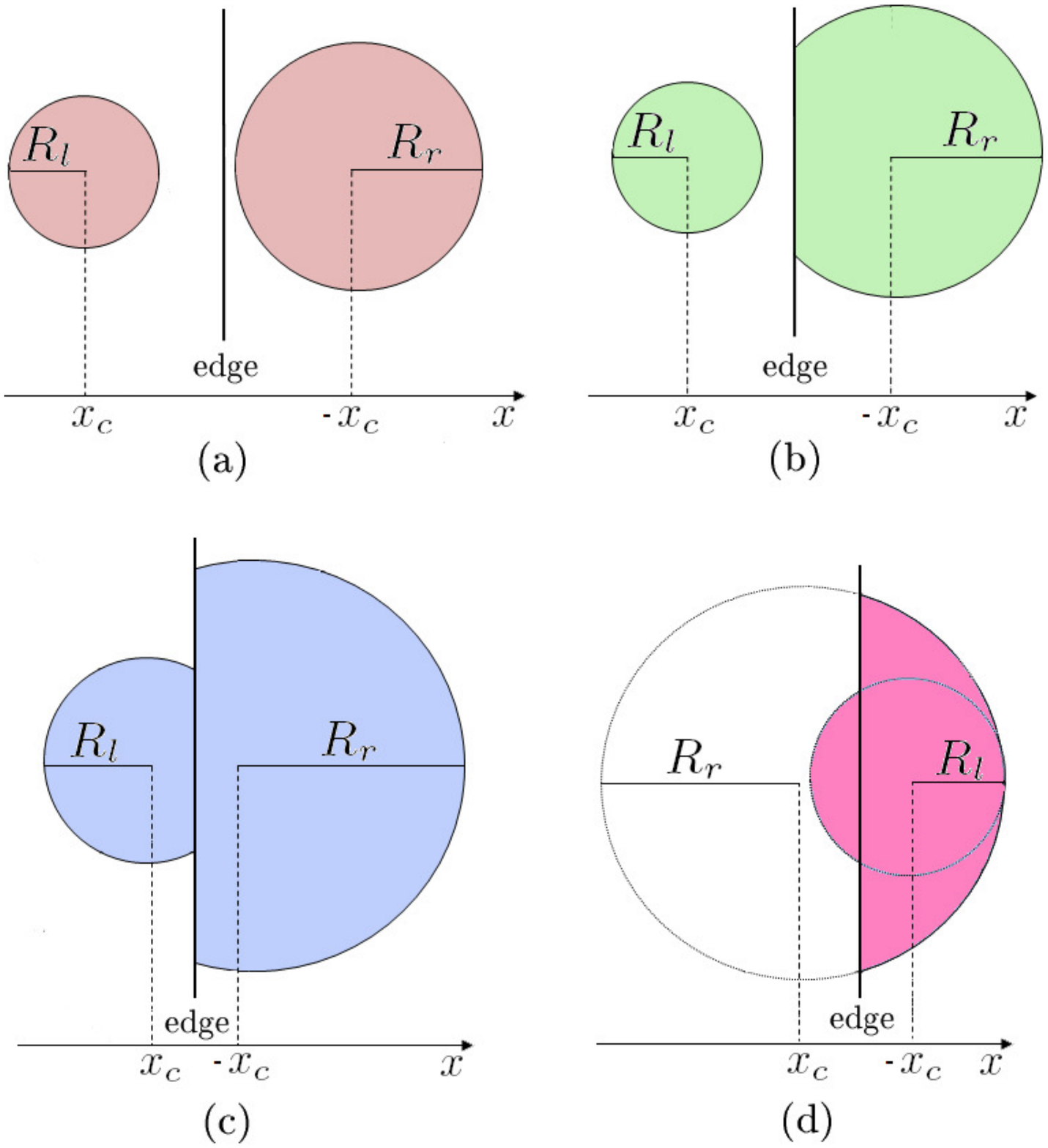}
	\caption{Classical cyclotron orbits for the "effective"  particle associated to the effective Hamiltonian, close to an armchair edge. (a) Each cyclotron orbit encloses an area ${\cal A}_l$ or ${\cal A}_r$, each of them quantized by the Bohr-Sommerfeld rule (\ref{bsq}). For an energy level $E_n=n$, one has $R_l=\sqrt{2 n -1}$ and $R_r=\sqrt{2 n +1}$. The ground state $n=0$ has no component on the left side. The associated spectrum is the bulk Landau levels spectrum displayed in the region $\cal C$ in Fig. \ref{fig:spectrearm}(b) When the left orbit approaches the edge, its image becomes a skipping orbit. This lifts the degeneracy between the energy levels as seen in Fig. \ref{fig:spectrearm} (region $\cal D$). (c) and (d) When $x_c$ is closer to the edge, both orbits are open skipping orbits. The quantization of the total area leads to the spectrum in region $\cal E$ of Fig.  \ref{fig:spectrearm}.}
		\label{fig:armso}
\end{figure}

For the zigzag edge, the two regions $\cal A$ and $\cal B$ in the spectrum (Fig. \ref{fig:zzhalperin}) can be associated to the two distinct cases represented in Fig. \ref{fig:zzso}.
\begin{eqnarray}
\mbox{(a) region ${\cal A}$,} \  && {\cal A}(E)= \pi R^2 = 2 \pi \left(n+\frac{1}{2}\right)    \\ 
\label{skipquant}
\mbox{(b) region ${\cal B}$,}  \  &&{\cal A}(E,x_c) = {R^2 \over 2} \left(2 \theta - \sin 2 \theta\right) = 2 \pi \left(n+\frac{3}{4}\right) \notag
\end{eqnarray}
from, which together with the relation $E=R^2/2 \pm 1/2$, we have deduced the energy levels, except near the region $x_c \simeq -R$.
For the armchair case, the situation is more involved, since we describe the classical motion of a massive particle {\em in the presence of its shifted image} (Fig. \ref{fig:armso}). Since the potential is asymmetric, we have defined two cyclotron radii related to the energy $E$ as $E= R_l^2/2 +1/2= R_r^2/2-1/2$. Now we have to consider three different cases corresponding to  the regions $\cal C$, $\cal D$ and $\cal E$ in Fig. \ref{fig:spectrearm}. 

 In the region $\cal C$ where $|x_c| > R_r > R_l$, the  orbit and its image are both closed cyclotron orbits. The quantization of their area~:
  \begin{eqnarray}
  \begin{split}
\mbox{left side:}  \qquad   & {\cal A}_l= \pi R_l^2 = 2 \pi \left(n_l+\frac{1}{2}\right)    \\
\mbox{right side:} \qquad   & {\cal A}_r = \pi R_r^2 = 2 \pi \left(n_r+\frac{1}{2}\right) 
\end{split}  
\end{eqnarray}
leads to the the energy levels  given by $E_n=n$, with $n=n_r=n_l+1$, what quantize the cyclotron radii as $R_r=\sqrt{2n +1}$ and $R_r=\sqrt{2n-1}$, the latest does not allow the $n=0$ level.

In the region $\cal D$ where $ R_r > |x_c| >  R_l$, the  left orbit is still closed but its image is  a skipping orbit. Keeping the notation $n=n_r=n_l+1$, we have~:
   \begin{eqnarray}
   \begin{array}{rlr}
\mbox{left side:}    & {\cal A}_l= \pi R_l^2 = 2 \pi (n-\frac{1}{2})   & n >0   \\[3mm]
\mbox{right side:}    &{\cal A}_r ={R_r^2 } (\theta - \frac{1}{2} \sin 2 \theta) = 2 \pi (n+\frac{1}{2})  & n \geq 0 
\end{array}
\end{eqnarray}
from which we obtain the energy levels in region $\cal D$. The degeneracy lift is classically explained by the fact that  both orbits have a different structure. The left orbit is closed and its semiclassical energy is still $E_n=n+1$, while the energy associated with the right skipping trajectory increases with $x_c$. Note that it is in this peculiar region that an inversion of the slope occurs.

  In the region $\cal E$ where $ R_r >   R_l > |x_c|$, both orbits are open and the area to be quantized is the total area
     \be
 {\cal A}_l+ {\cal A}_r =  2 \pi \left(n'+\frac{1}{2}\right)   \qquad n' \geq 0 
    \ee
Since the action has almost doubled from region $\cal D$ to region $\cal E$, the resulting  energy spectrum
is twice denser, and the two-fold degeneracy of the levels has been removed.

																							 \section{Two edges}

\begin{figure}[!ht]
	\centering
		  \includegraphics[width=9cm]{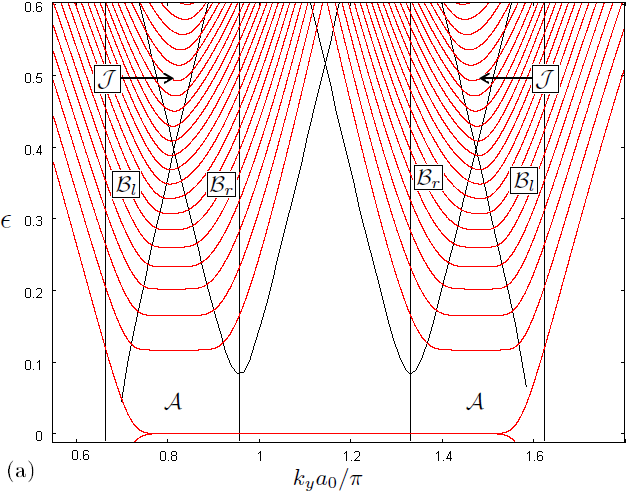}
	\centering
		\includegraphics[width=9cm]{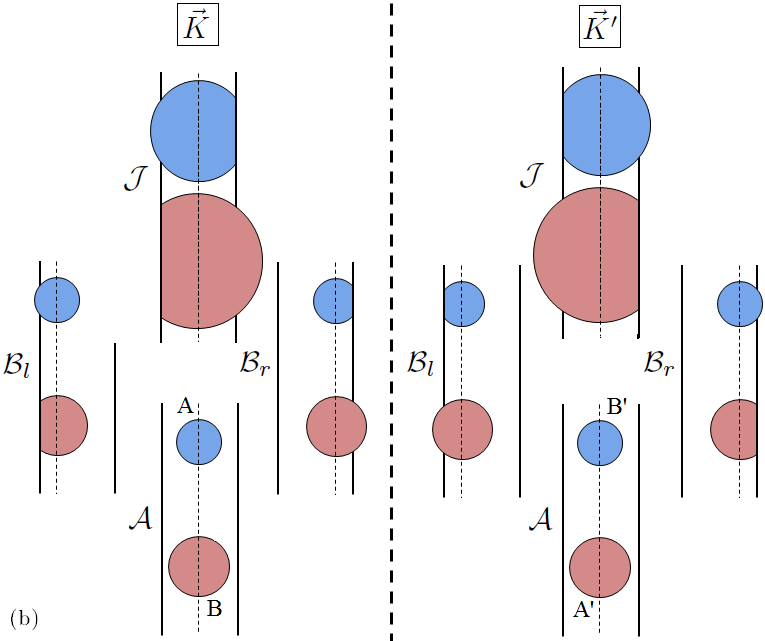}		  
	\caption{(a) Red curves: Low energy spectrum for a zigzag ribbon in a magnetic field in units of $t$. The width $L$ of the ribbon is $L= 174 a$ and the dimensionless magnetic flux $f$ is $f=0.00126$, so that $L/\ell_B= 9.6$. The black curves indicate the situations where the classical cyclotron orbits touch an edge.   (b) Schematic representation of the different situations for the position of the cyclotron orbits with respect to the edges of the ribbon, as discussed in the text.}
	\label{fig:ZZ2edges}
\end{figure}

\begin{figure}[!ht]
	\centering
		\includegraphics[width=9cm]{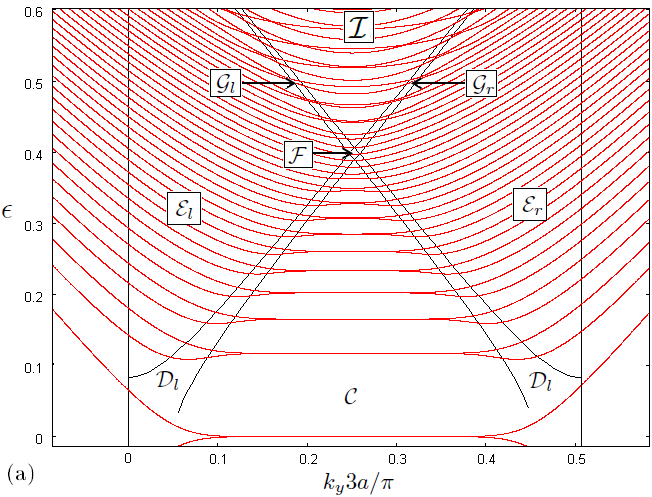}
			\centering
		\includegraphics[width=9cm]{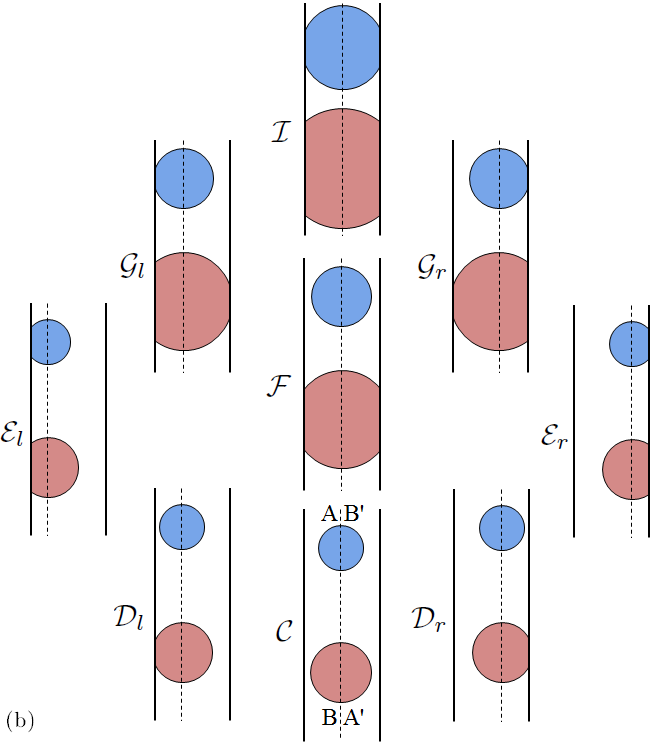}
	\caption{(a) Red curves: Low energy spectrum for an armchair ribbon in a magnetic field in units of $t$. The width $L$ of the ribbon is $L= 201 a_0/2$ and the dimensionless magnetic flux $f$ is $f=0.00126$, so that $L/\ell_B= 9.6$. The black curves indicate the situations where the classical cyclotron orbits touch an edge.  (b) Schematic representation of the different situations for the position of the cyclotron orbits with respect to the edges of the ribbon.}
	\label{fig:AC2edges}
\end{figure}

Up to now, we have considered the evolution of the spectrum in the vicinity of one edge. If the two edges are sufficiently far apart compared to the cyclotron length $\ell_B$, the spectrum  can be treated independently on both sides. We now consider the case of a narrow ribbon, whose width $L$ is of the order of a few magnetic lengths. The low energy spectrum in a ribbon of width $L=9.6 \ell_B$ is shown on Fig. \ref{fig:ZZ2edges}(a) and \ref{fig:AC2edges}(b), respectively for zigzag and armchair ribbons.

In the zigzag case, in each valley, the spectrum clearly exhibits three different regions, which correspond to the geometry of the orbits depicted in Fig. \ref{fig:ZZ2edges}(b) . Regions ${\cal A}$ corresponds to bulk cyclotron orbits. Regions ${\cal B}_l$ and ${\cal B}_r$ have been already discussed and correspond to a skipping orbit along a single boundary (left or right). Each sublattice is characterized by a cyclotron orbit (in $K$ valley, small blue orbit for $A$ sites and large pink orbit for $B$ sites; the opposite for the $K'$ valley).  As explained in the text, in $K$ valley, the $A$ cyclotron orbits sees only the right edge and the $B$ cyclotron orbits sees only the left edge (the opposite in the $K'$ valley). At high energy, the new region ${\cal J}$ corresponds to the situation where both cyclotron orbits intersect the two boundaries. The black curves correspond to the situations where a cyclotron orbit precisely touches a boundary. Their equation is
\be 
\ep(k_y)=t \frac{\sqrt{3}a_0}{2}\sqrt{(q_y-\Delta q_y)^2 \pm 1/\ell_B^2} 
 \label{graze}
\ee
where $\Delta q_y=0$ for the left edge and $\Delta q_y=L/\ell_B^2$ for the right edge. By a simple quantization of the areas shown in Fig. \ref{fig:ZZ2edges}(b),  one can obtain the full low energy spectrum in a very good approximation, excepted when the orbits graze the boundaries, that is in the vicinity of the black curves. It is obvious on this figure that the two edges do not play exactly the same role, and that, for a given valley, the spectrum is not exactly symmetric. Furthermore, in each valley, an additional dissymetry is seen at high energy because the linear approximation of the low energy Hamiltonian breaks (\ref{DiracMagn}) down.

The case of armchair boundary conditions is more involved. In Fig. \ref{fig:AC2edges}(a), one can distinguish 9 regions, which corresponds to the geometries of the classical orbits depicted in Fig. \ref{fig:AC2edges}(b). The black curves correspond to the situations where a cyclotron orbit precisely touches a boundary. Their equation is given by Eq. (\ref{graze}). Region ${\cal C}$ corresponds to bulk cyclotron orbits. The blue  cyclotron orbits are related to the $A$ and $B'$ eigenfunctions, whereas the pink cyclotron orbits are related to the $A'$ and $B$ eigenfunctions.  Regions   ${\cal E}_l$  and   ${\cal E}_r$  have been already discussed and correspond to  skipping orbits along a single boundary (left or right). In the intermediate regions ${\cal D}_l$  and   ${\cal D}_r$, only one of the two orbits touches the edge. At high energy, in the region ${\cal I}$,  the two cyclotron orbits touch the two edges of the ribbon. The more exotic regions ${\cal F}$, ${\cal G}_l$ and  ${\cal G}_r$ have a simple geometric interpretation in Fig.  \ref{fig:AC2edges}(b).
\section{Conclusion}
\label{sec:conclu}

We have investigated the spectra of graphene ribbons in a magnetic field with zigzag and armchair boundary conditions. We have first revisited these spectra numerically and revealed a remarkable structure in the repartition of the edge states in zigzag ribbons, that we explain in this paper. We notice that these remarkable behaviours  of the edge states could be observed by scanning tunneling microscopy (STM) or spectroscopy (STS) techniques.\cite{abanin,matsui,niimi,niimi2}

Next we have described and calculated these edge states with simple analytic tools. For both types of ribbons, effective Schr\"odinger equations with a specific double well potential have been derived at low energy. This potential is asymmetric in energy when the valleys $\K$ and $\Kp$ are coupled (armchair edge) and symmetric otherwise (zigzag edge). The eigenenergies of these potentials have been calculated recently.\cite{abanin} Another recent work\cite{rakyta} provided a semiclassical framework to study analytically the edge states for both zigzag and armchair edges, but, this approach does not furnish the full low energy spectrum. We have developped here two semiclassical methods to calculate the energy edge states analytically. A very simple one consists in using the Bohr-Sommerfeld quantization of the action related to the effective Schr\"odinger equations. This approximation captures  the essential of the physical picture except when the cyclotron radius of the effective particule is of the order of the distance to the edge. This approach also reveals different regions as a function of the energy and the distance to the edge, where the skipping orbits are quantized in different ways.   The second method is based on the WKB formalism, and accounts for the overlap of the wave function when it is close to the edge. Consequently, this more sophisticated analytical approach perfectly describes the edge states whatever the distance of the center of the cyclotron motion to the edge. In particular, it quantitatively describes an interesting region in the armchair case (called region $\cal D$ in the paper) where the energy does not increase monotonously. This implies that the drift velocity along the edge $v^n_{drift}=\partial \ep_n/\partial k_y\propto \partial \ep_n/\partial x_c$ may change in sign when the position $x_c$ varies. These WKB results perfectly fit  numerical exact calculations.   
\\
\\

	 \textit{Acknowledgments }-  We  acknowledge useful discussions with J.-N Fuchs and  M.-O. Goerbig.	This work is supported by the NANOSIM-GRAPHENE project (ANR-09-NANO-016-01) of ANR/P3N2009.

																	\appendix*
																					\section{The WKB method}
\emph{In the whole appendix, we define the classical action as half of the one defined in (\ref{action}). }\\

In this appendix, we present the detailed calculations on the energy levels $E_n(x_c)$ within the WKB approximation, for the potential $V(x)$ defined as~: 
 \begin{eqnarray}
 V(x)=
 \begin{cases}
 &\frac{1}{2}(x-x_c)^2+\frac{V_0}{2} \ \ \ \textrm{for} \ \ x \ \leq 0\\
      &\frac{1}{2}(x+ x_c)^2-\frac{V_0}{2} \ \ \ \textrm{for} \ \ x \ \geq 0
 \end{cases}
 \end{eqnarray}
 The case $V_0=0$ has been introduced in section \ref{sec:zz} for the study of edge states in zigzag nanoribbons, while the case $V_0=1$ has been used in section \ref{secarmchair} to describe armchair ribbons. 
\begin{figure}[!ht]
	\centering
		\includegraphics[width=8cm]{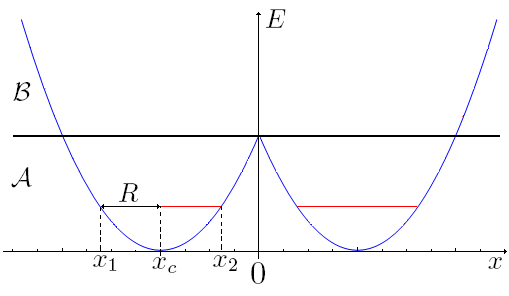}
		\caption{The double symmetric well. The two wells are centered in $\pm x_c$. The two turning points in the left well are $x_1$ and $x_2$.}
	\label{fig:puisym1}
\end{figure}

																									\subsection{Double symmetric harmonic well}
	
 We consider the peculiar case $V_0=0$, that is  a double symmetric harmonic potential~:
 \begin{equation}
 V_{Sym}(x)=
\frac{1}{2}(|x|+x_c)^2 \
 \end{equation}
 For a given energy $E$, we define the cyclotron radius as $R =\sqrt{2 E}$ (Fig. \ref{fig:puisym1}). In addition, in the whole appendix, the factor $2$ in the definition of the action (\ref{action}) is missing: the actions $S_l$ and $S_r$  are now defined as the half of those introduced in Sec. \ref{sec:sc}.

																									\subsubsection{Region $\cal A$~: $R\leq\left|x_c\right|$}
																		               \label{sec:symwellle}
																		
This case, characterized by $E \leq x_c^2/2$, corresponds to  the region $\cal A$  in Fig.  \ref{fig:puisym1}.  We write the WKB wave functions in both wells and then the matching conditions in $x=0$. As the potential is symmetric with $x$, we   focus on the left well ($x<0$). The energy $E$ defines two classical turning points  at positions $x_1=x_c -R$ and $x_l=x_c+R$ (Fig. \ref{fig:puisym1}). Within  the WKB approximation, the connection procedure near $x_1$ implies that the wave function in the well reads~:
\begin{eqnarray}
\varphi_l^\leftarrow(x)&=& \frac{C_l}{\sqrt{k(x)}}\sin({\cal S}(x_1,x)+\frac{\pi}{4})
\label{wkb}
\end{eqnarray}
where the left arrow indicates that this wave function matches the correct connection procedure near the left turning point $x_1$. $C_l$ is a constant and ${\cal S}(x_1,x)$ is the partial action in the left well~:
\be
{\cal S}(x_1,x)= \int_{x_1}^{x}dx\sqrt{E-V_{Sym}(x)}  \ . \label{Sx1x}
\ee																																					
The above expression (\ref{wkb}) breaks down near the second  turning point $x_l=x_c+R$. Near this point we linearize the potential as    $V_{Sym}(x)\approx \frac{R^2}{2}+R(x-x_l)$ and the  Schr\"odinger equation reads~:
\begin{equation}
\left(\frac{d^2}{d x^2}-2 R (x-x_l)\right)\varphi^l(x)=0.
\end{equation}
The solution of this equation is given by a combination of Airy functions~:
\begin{eqnarray}
\label{wkbright}
&&\varphi_l^\rightarrow(x)=\bar{\alpha_l} \Ai(ax+x_0)+\bar{\beta_l} \Bi(ax+x_0) \label{AiBi1} \\
&&a^3\equiv2 R\\
&&x_0\equiv(2R)^{1/3}\left(R-\left|x_c\right|\right)
\end{eqnarray}
where the right arrow indicates that this wave function must obey proper matching conditions near the right turning point $x_l$. $\bar{\alpha_l}$ and $\bar{\beta_l}$ are constants. Inside the well, this wave function has the
  asymptotic   expansion~:
\begin{eqnarray}
\varphi_l^\rightarrow(x) \simeq  \frac{\bar{\alpha_l}}{\sqrt{\pi}\left| z \right|^{1/4}} \sin\left(\frac{2}{3}\left|z\right|^{3/2} +\frac{\pi}{4}\right) \notag \\
  +  \frac{\bar{\beta_l}}{\sqrt{\pi}\left| z \right|^{1/4}} \cos\left(\frac{2}{3}\left|z\right|^{3/2}  +\frac{\pi}{4}\right)
\label{asymp}
\end{eqnarray}
\begin{eqnarray}
\text{with}\ \ \left|z\right|\equiv-x_0-ax=\frac{k^2(x)}{a^2}  \ .
\label{z}
\end{eqnarray}
The argument in the trigonometric functions can be related to the partial action between $x$ and $x_l$
\begin{eqnarray}
\frac{2}{3}\left|z\right|^{3/2}= \int_x^{x_l}dx\sqrt{E-V_{Sym}(x)}= S_l-{\cal S}(x_1,x)
\label{S}
\end{eqnarray}
where $S_l$ is the total action in the left well. Therefore $\varphi_l^\rightarrow(x)$ can be rewritten in the form~:
\begin{eqnarray}
\varphi_l^\rightarrow(x)  \simeq  \frac{\alpha_l}{\sqrt{k(x)}}\cos\left({\cal S}(x_1,x)+\frac{\pi}{4}-S_l\right)\notag \\
+\frac{\beta_l}{\sqrt{k(x)}}\sin\left({\cal S}(x_1,x)+\frac{\pi}{4}- S_l\right)
\label{asymp2}
\end{eqnarray}
with $\alpha_l=\bar{\alpha}_l \sqrt{\frac{a}{\pi}}$ (idem for $\beta_l$). Then, we impose $\varphi_l^\leftarrow(x)=\varphi_l^\rightarrow(x)$  inside the well. This implies  $\alpha_l=C_l\sin S_l$ and $\beta_l=C_l\cos S_l$ what finally leads to the important relation~:
\begin{equation}
\tan S_l=\frac{\bar{\alpha_l}}{\bar{\beta_l}}=\frac{\alpha_l}{\beta_l}.
\label{tanl}
\end{equation}
As the two wells are identical, we obtain the same relation for the right well~:
\begin{equation}
\tan S_r=\frac{\bar{\alpha_r}}{\bar{\beta_r}}=\frac{\alpha_r}{\beta_r}
\label{tanr1}
\end{equation}
and, similarly to (\ref{AiBi1}), the wave function in the right well reads near $x=0$~:
\begin{eqnarray}
&&\varphi_r^\leftarrow(x)=\bar{\alpha_r} \Ai(-ax+x_0)+\bar{\beta_r} \Bi(-ax+x_0).
\end{eqnarray}																			
The next step is to impose  the matching  of the two wave functions  and their derivatives in $x=0$~:
\begin{eqnarray}
&&\varphi_l^\rightarrow(0)= \varphi_r^\leftarrow(0)  \ \ \ , \ \ \
 \partial_x\varphi_l^\rightarrow (0)= \partial_x\varphi_r^\leftarrow  (0)
\end{eqnarray}
These matching conditions give the two equations~:
\begin{eqnarray*}
\bar{\alpha_l} \Ai(x_0)+ \bar{\beta_l} \Bi(x_0) &=& \bar{\alpha_r} \Ai(x_0)+ \bar{\beta_r} \Bi(x_0)\\
\bar{\alpha_l} \Ai'(x_0)+ \bar{\beta_l} \Bi'(x_0) &=& -\bar{\alpha_r} \Ai'(x_0)- \bar{\beta_r} \Bi'(x_0)
\end{eqnarray*}
from which the   ratio $\frac{\bar{\alpha_r}}{\bar{\beta_r}}$ is extracted as~:
\begin{equation}
\frac{\bar{\alpha_r}}{\bar{\beta_r}}=-\frac{\frac{\bar{\alpha_l}}{\bar{\beta_l}}\left(\Bi'(x_0) \Ai(x_0) +\Bi(x_0)\Ai'(x_0)\right)+ 2\Bi'(x_0)\Bi(x_0)}{ \frac{\bar{\alpha_l}}{\bar{\beta_l}}2\Ai'(x_0)\Ai(x_0)+\Ai'(x_0)\Bi(x_0)+\Ai(x_0)\Bi'(x_0)}.
\label{alphabeta}
\end{equation}
The symmetry of the potential implies  $S_l=S_r$, so that, from  (\ref{tanl}) and (\ref{tanr1}), we have $\frac{\bar{\alpha_r}}{\bar{\beta_r}}=\frac{\bar{\alpha_l}}{\bar{\beta_l}}\equiv X$. The relation (\ref{alphabeta}) becomes a simple polynomial   for the unknown quantity $X$~:
\begin{eqnarray}
\Ai'(x_0) \Ai(x_0) X^2 &+& \left(  \Ai'(x_0)\Bi(x_0)+\Ai(x_0)\Bi'(x_0) \notag \right)X\\
&+& \Bi(x_0) \Bi'(x_0)=0
\label{poly1}
\end{eqnarray}
whose solutions
\begin{eqnarray}
X^{S}=-\frac{\Bi'(x_0)}{\Ai'(x_0)} \ \ \  \ \ \ X^{AS}=-\frac{\Bi(x_0)}{\Ai(x_0)}
\end{eqnarray}
correspond respectively to the symmetric and antisymmetric wave functions. The action $S_l$ is quantized by the condition $\tan S_l= X^{S,AS}$, so that
 $S_l= \pi (n + \gamma)$ with $0 \leq \gamma < 1$, with
two solutions for $\gamma$~:
\begin{eqnarray}
\gamma^{S}(E,x_c)&=& - \frac{1}{\pi} \arctan \frac{\Bi'(x_0)}{\Ai'(x_0)} \\
\label{miles}
\gamma^{AS}(E,x_c)&=& 1 - \frac{1}{\pi} \arctan \frac{\Bi(x_0)}{\Ai(x_0)}
\label{mileas}
\end{eqnarray}
In region ${\cal A}$, the action is very simply related to the energy, $S=l = \pi R^2/2= \pi E$, so that from (\ref{miles},\ref{mileas}), we obtain the implicit equations~:
\begin{eqnarray}
\label{eqimplle1}
E_n^{S}=n + \frac{1}{\pi}\arctan\left(-\frac{\Bi'(x_0)}{\Ai'(x_0)}\right) \\
E_n^{AS}=n+1 +\frac{1}{\pi}\arctan\left(-\frac{\Bi(x_0)}{\Ai(x_0)}\right) \ .
\label{eqimplle2}
\end{eqnarray}
where $x_0$ itself depends on the energy~:
\begin{eqnarray*}
x_0&=&\left(2\sqrt{2 E_n}\right)^{1/3}(\sqrt{2 E_n}-\left|x_c\right|)
\end{eqnarray*}
These implicit equations are solved numerically and we obtain the spectrum (region ${\cal A}$) shown in  Fig. \ref{fig:spectre}.


																					\subsubsection{Region $\cal B$~: $\left|x_c\right|\leq R$}
										                       \label{sec:symwellhe}
This is  the region $\cal B$ illustrated in Fig.  \ref{fig:puisym1}. The expression of the WKB wave function matching the correct connection procedure near $x_1$ is still given by Eq. (\ref{wkb}). The second turning point in $x_2$ does not exist anymore, and we have to know the expression of the wave function near $x=0$. It reads~:
 \begin{eqnarray}
\varphi_l^\rightarrow(x) &=&\bar{\alpha_l} \Ai(y_0+ax)+\bar{\beta_l} \Bi(y_0+ax)\\
a^3& \equiv &2\left|x_c\right|\\
y_0& \equiv &\frac{x_c^2-R^2}{\left(2 \left| x_c \right|\right)^{2/3}}
\label{y0}
\end{eqnarray}
Far from $x=0$, the Airy functions can be expanded to obtain~:
\begin{eqnarray*}
\varphi_l^\rightarrow(x) \simeq    \frac{\bar{\alpha_l}}{\sqrt{\pi}\left| z \right|^{1/4}} \sin\left(\frac{2}{3}\left|z\right|^{3/2} +\frac{\pi}{4}\right) \notag \\  +  \frac{\bar{\beta_l}}{\sqrt{\pi}\left| z \right|^{1/4}} \cos\left(\frac{2}{3}\left|z\right|^{3/2}  +\frac{\pi}{4}\right)
\end{eqnarray*}
where we have set~:
\begin{eqnarray}
\left|z\right|\equiv-y_0-ax=\frac{k^2(x)}{a^2}\ \ \text{,}\ \ \alpha_l \equiv \frac{\bar{\alpha_l}}{\sqrt{\pi}}(2 \left|x_c  \right|)^{1/6}.
\end{eqnarray}
The arguments in the trigonometric functions can be related to the partial action between $x$ and $0$~:
\begin{eqnarray}
\frac{2}{3}\left| z \right|^{3/2}&=&\int_x^0{dx \sqrt{R^2-x_c^2-2\left|x_c\right|x }}+\delta \notag \\
                                 &=& S_l -{\cal S}(x_1,x)+ \delta 
\ee
where the quantity~: 
\be                                 
\delta & \equiv & \frac{(R^2-x_c^2)^{3/2}}{3 \left|x_c\right|}
\end{eqnarray}
accounts for the step between the potential $V(x=0)$ and the energy $E$. The wave function  $\varphi_l^\rightarrow(x)$ can thus be rewritten as~:
\begin{eqnarray}
\varphi_l^\rightarrow(x)  \simeq  \frac{\alpha_l}{\sqrt{k(x)}}\cos({\cal S}(x_1,x)+\frac{\pi}{4}-S_l-\delta)\notag \\
+\frac{\beta_l}{\sqrt{k(x)}}\sin({\cal S}(x_1,x)+\frac{\pi}{4}- S_l-\delta) \ .
\label{asymp2}
\end{eqnarray}
The matching of the two wave functions $\varphi_l^\leftarrow(x)=\varphi_l^\rightarrow(x)$ gives~:
\begin{eqnarray}
\label{slhe}
&&\tan(S_{l}+\delta)=  \frac{\alpha_{l}}{\beta_{l}}
\end{eqnarray}
and a similar expression for the right well. Again we impose the continuity of the wave function and its derivative in $x=0$ and obtain the two equations~:
\begin{eqnarray*}
\bar{\alpha_l} \Ai(y_0)+ \bar{\beta_l} \Bi(y_0) &=& \bar{\alpha_r} \Ai(y_0)+ \bar{\beta_r} \Bi(y_0)\\
\bar{\alpha_l} \Ai'(y_0)+ \bar{\beta_l} \Bi'(y_0) &=& -\bar{\alpha_r} \Ai'(y_0)- \bar{\beta_r} \Bi'(y_0)
\end{eqnarray*}
from where we extract the ratio $\alpha_r/\beta_r$. Then, using the symmetry $\alpha_{l}=\alpha_{r}$ and $\beta_{l}=\beta_{r}$, we obtain a polynomial in $X\equiv\alpha_l/\beta_l$~:
\begin{eqnarray}
\Ai'(y_0) \Ai(y_0) X^2 &+& \left(  \Ai'(y_0)\Bi(y_0)+\Ai(y_0)\Bi'(y_0)  \right)X \notag \\
&+& \Bi(y_0) \Bi'(y_0)=0 \ .
\end{eqnarray}
The only difference with the polynomial (\ref{poly1}) for the region ${\cal A}$  consists in the substitution $x_0\rightarrow y_0$. Consequently, the solutions are~:
\begin{eqnarray}
\label{xhe}
X^{S}=-\frac{\Bi'(y_0)}{\Ai'(y_0)} \ \ \ \ \ \
X^{AS}=-\frac{\Bi(y_0)}{\Ai(y_0)} \ .
\end{eqnarray}
The action $S_l$ is quantized by the condition $\tan S_l= X^{S,AS}$, so that
 $S_l= \pi (n + \gamma)$ with $0 \leq \gamma < 1$, with
  two solutions for $\gamma$~:
\begin{eqnarray}
\gamma^{S,AS}(E,x_c)&=& \sigma\left[\frac{1}{\pi}( \arctan X^{S,AS} - \delta )\right]
\label{mile2}
\end{eqnarray}
where $\sigma(x)=x - [x]$,  $[x]$ being the   next smallest   integer.  By construction $0 \leq \sigma[x] <1$. 
 
\begin{figure}[!ht]
	\centering
		\includegraphics[width=9cm]{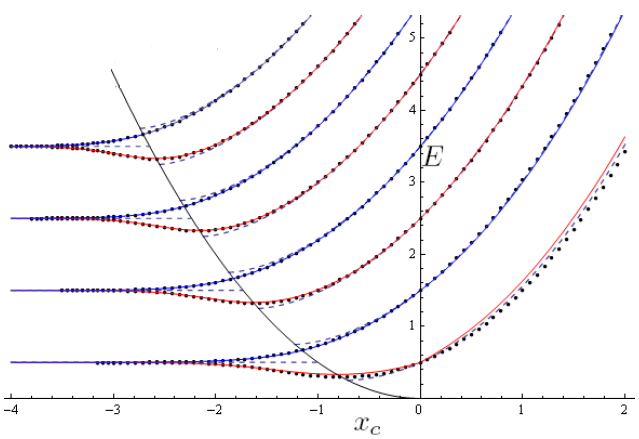}
	\caption{Semiclassical spectra $E_n(x_c)$ within the semiclassical approximation (Dashed lines) $\gamma=cst$, (full lines) WKB method. The dots are the exact numerical solutions of the Schr\"odinger equation with the potentiel $V_{sym}(x)$. The two regions are separated by the parabola $E=x_c^2/2$. For the zigzag problem, we keep only antysymmetric states that are the high energy levels (blue lines).}
		\label{fig:spectre}
\end{figure}

\begin{figure}[!ht]
	\centering
		\includegraphics[width=9cm]{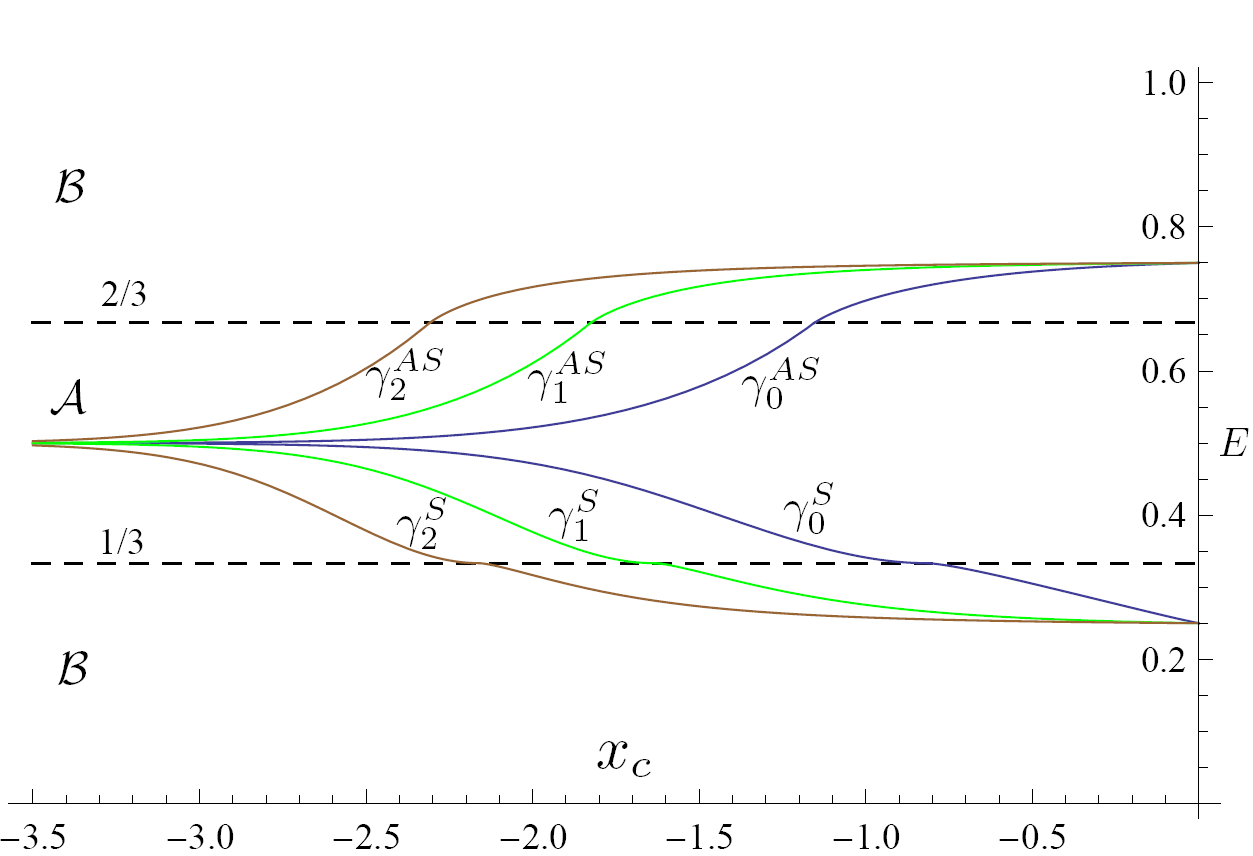}
			\caption{Mismatch index $\gamma^{S,AS}(x_c)$ as a function of $x_c$, for (blue) $n=0$, (green) $n=1$ and (brown) $n=2$. For $x_c=R$, $\gamma^{S}(R)=1/3$ whereas $\gamma^{AS}(R)=2/3$ (horizontal dashed lines). For $x_c=0$, $\gamma^{S}(0)=1/4$ and $\gamma^{AS}(0)=3/4$.}
		\label{fig:mi}
\end{figure}

The last step is to relate the energy to the action. This relation is not linear as in region ${\cal A}$ but it now reads (see Eq. \ref{action2})~: 
\be
S_l= E \left(\theta - {1 \over 2} \sin 2 \theta\right)  
\ee
where $\theta= \arccos x_c / R= \arccos x_c/\sqrt{2E}$ has the meaning of a angle (Sec. \ref{secarea}). From this equation, together with the quantization condition~:
\be
 S_l= \pi (n + \gamma^{S,AS}) 
\ee
where $\gamma^{S,AS}(E,x_c)$ are functions of $E$ and $x_c$ through Eqs. (\ref{mile2}, \ref{y0}), we can extract   numerically the   eigenenergies $E_n^S$ and $E_n^{AS}$   and plot them  as a function of $x_c$ in Fig. \ref{fig:spectre}. In Fig. (\ref{fig:mi}) we have also plotted the $x_c$ dependence   of the mismatch index $\gamma^{S,AS}(x_c)$.

																		\subsection{Double harmonic asymmetric well}
\begin{figure}[!ht]
	\centering
		\includegraphics[width=9cm]{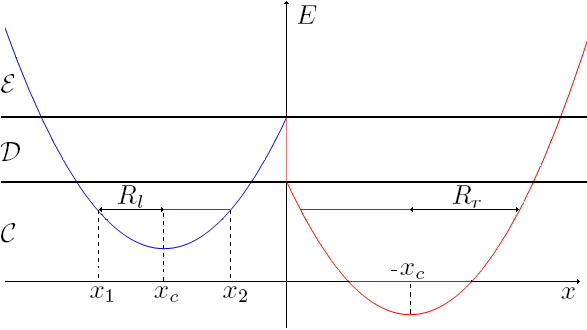}
			\caption{Double asymmetric well $V_{Asym}(x)$: $\cal C$, $\cal D$ and $\cal E$ refer to the three regions discussed in the text. For a given energy $E$, we define two cyclotron radii $R_l$ and $R_r$.}
	\label{fig:potential2}
\end{figure}

In this section  we introduce an asymmetry $V_0=1$ in the potential which becomes~:
\begin{eqnarray}
 V_{Asym}(x)=
 \begin{cases}
 &\frac{1}{2}(x-x_c)^2+\frac{1}{2} \ \ \ \textrm{for} \ \ x \ \leq 0\\
      &\frac{1}{2}(x+ x_c)^2-\frac{1}{2} \ \ \ \textrm{for} \ \ x \ \geq 0
 \end{cases}
 \end{eqnarray}
and calculate   the spectrum $E_n(x_c)$ within the WKB approximation. We now introduce two cyclotron radii $R_l$ and $R_r$. The relation between these parameters and the classical energy is given by~:
 \begin{eqnarray}
E=\frac{R_l^2+1}{2}=\frac{R_r^2-1}{2}.
 \end{eqnarray}
We now have to  distinguish three regions delimited by $\left|x_c\right|=R_l$ and $\left|x_c\right|=R_r$ and illustrated in Fig. \ref{fig:potential2}. 
The total action in these three regions reads~:
\begin{equation}
\begin{split}
\textrm{Region  {\cal C}} &\qquad  S_t=\frac{\pi}{2}R_l^2+\frac{\pi}{2}R_r^2\\
\textrm{Region {\cal D}} &\qquad  S_t=\frac{\pi}{2}R_l^2  +   \frac{R_r^2}{2} [\theta_r -\frac{1}{2}\sin(2 \theta_ r)  ] \\
\textrm{Region {\cal D}} &\qquad  S_t=\frac{R_l^2}{2} [\theta_l -\frac{1}{2}\sin(2 \theta_ l)  ]  +   \frac{R_r^2}{2} [\theta_r -\frac{1}{2}\sin(2 \theta_ r)  ] 
\end{split}
\end{equation}
We treat the three regions separately.

																		\subsection{Region $\cal C$~: $R_l \leq R_r\leq \left|x_c\right|$}
																		
This  region is explicited  in Fig. \ref{fig:potential2}. Because of the asymmetry of the potential, the action is now different in each well~:
\begin{eqnarray}
S_l&=&\frac{\pi}{2}R_l^2={\pi}\left(E -1/2\right)\\
S_r&=&\frac{\pi}{2}R_r^2=\pi\left(E+1/2\right).
\label{sas1}
\end{eqnarray}
The WKB wave function has the same form as in  Eqs. (\ref{wkb}), (\ref{wkbright}) and (\ref{asymp}) written for the symmetric potential so that we still have the relations (\ref{tanl}) and (\ref{tanr1}). Since the potential is now asymmetric,   the expression of the  wave function near the two inner turning points located in $-\left|x_c\right|+R_l$ and   $\left|x_c\right|-R_r$ is now~:
\begin{eqnarray}
\varphi_l^\rightarrow(x)&=&\bar{\alpha}_l \Ai(a_lx+x_l)+\bar{\beta}_l \Bi(a_lx+x_l)\\
\varphi_r^\leftarrow(x))&=&\bar{\alpha}_r \Ai(-a_rx+x_r)+\bar{\beta}_l \Bi(-a_rx+x_r)
\end{eqnarray}
with
\begin{eqnarray}
a_l^3&\equiv&2R_l  \ \ \ \ \ \   \   a_r^3\equiv2R_r\\
x_l & \equiv & (2 R_l)^{1/3}\left(R_l-\left|x_c\right|\right)\\
x_r & \equiv & (2 R_r)^{1/3}\left(R_r-\left|x_c\right|\right) \ .
\end{eqnarray}
Then, by imposing the current conservation and the continuity of these wave functions in $x=0$, we obtain the two equations~:
\begin{eqnarray}
\bar{\alpha}_l \Ai(x_l)+ \bar{\beta}_l \Bi(x_l) &=& \bar{\alpha}_r \Ai(x_r)+ \bar{\beta}_r \Bi(x_r)\\
\bar{\alpha_l} a_l \Ai'(x_l)+ \bar{\beta_l} a_l \Bi'(x_l) &=& -\bar{\alpha}_r a_r \Ai'(x_r)- \bar{\beta}_r a_r \Bi'(x_r)\notag
\end{eqnarray}
from where we extract the ratio $\frac{\bar{\alpha}_r}{\bar{\beta}_r}$ as~:
\begin{widetext}
\begin{equation}
\frac{\bar{\alpha}_r}{\bar{\beta}_r}=-\frac{ \frac{\bar{\alpha}_l}{\bar{\beta}_l}\left(\Bi'(x_r)\Ai(x_l)+\Bi(x_r)\Ai'(x_l)\left(\frac{R_l}{R_r}\right)^{1/3}\right)+ \Bi'(x_r)\Bi(x_l)+\Bi(x_r)\Bi'(x_l)\left(\frac{R_l}{R_r}\right)^{1/3}}{   \frac{\bar{\alpha}_l}{\bar{\beta}_l}\left(\Ai'(x_r)\Ai(x_l)+\Ai(x_r)\Ai'(x_l)\left(\frac{R_l}{R_r}\right)^{1/3}\right)  +\Ai'(x_r)\Bi(x_l)+\Ai(x_r)\Bi'(x_l)\left(\frac{R_l}{R_r}\right)^{1/3}}   \   .
\label{alphabetaas1}
\end{equation}
\end{widetext}
Since $S_l \neq S_r$, the ratios $\frac{\bar{\alpha}_r}{\bar{\beta}_r}$ and $\frac{\bar{\alpha}_l}{\bar{\beta}_l}$ are now different. Introducing the total action $S_t=S_r+S_l=2 \pi E$, these coefficients are related as~:
\begin{equation}
\frac{\alpha_r}{\beta_r}=\frac{\tan{S_t}-\alpha_l/\beta_l}{1+\alpha_l/\beta_l \tan{S_t}}  \ .
\label{tan}
\end{equation}
 Then, we insert the relation (\ref{tan}) into (\ref{alphabetaas1}) to obtain a polynomial in $X_l\equiv\alpha_l/\beta_l$ whose the coefficients are only functions of $E$ and $x_c$~:
\begin{widetext}
\begin{eqnarray}
\label{polas1}
&&\left\{\tan{S_t}\left(\Bi'(x_r)\Ai(x_l)+\left(\frac{R_l}{R_r}\right)^{1/3}\Bi(x_r)\Ai'(x_l)\right)-\Ai'(x_r)\Ai(x_l)-\left(\frac{R_l}{R_r}\right)^{1/3}\Ai'(x_l)\Ai(x_r)\right\}   X^2_l\\
&&\left\{\tan{S_t}\left(\Ai'(x_r)\Ai(x_l)+\Bi'(x_r)\Bi(x_l) +  \left(\frac{R_l}{R_r}\right)^{1/3}\left[\Ai'(x_l)\Ai(x_r)+\Bi'(x_r)\Bi(x_l)\right]\right)
\right.
  \nonumber\\
&& \hspace{2.8cm}
 \left.
+\Bi'(x_r)\Ai(x_l)-\Ai'(x_r)\Bi(x_l)+\left(\frac{R_l}{R_r}\right)^{1/3}\left[  \Bi(x_r)\Ai'(x_l)-\Ai(x_r)\Bi'(x_l) \right]\right\}X_l\nonumber\\
&&\tan{S_t}\left(\Ai'(x_r)\Bi(x_l)+\left(\frac{R_l}{R_r}\right)^{1/3}\Ai(x_r)\Bi'(x_l)\right)+\Bi'(x_r)\Bi(x_l)+\left(\frac{R_l}{R_r}\right)^{1/3}\Bi(x_r)\Bi'(x_l)=0 \nonumber
\end{eqnarray}
\end{widetext}
with $S_t=2 \pi E$. Because of the asymmetry of the potential, the coefficients of the polynomial are much more complicated than for the symmetric case (Eq. \ref{poly1}). We check that for $R_r=R_l$ and $x_l=x_r=x_0$ we recover   the symmetric case (Eq. \ref{poly1}). The polynomial (\ref{polas1}) has still two solutions $X_l^S$ and $X_l^{AS}$, from which we obtain the action $S_l$. It has the from $S_l=\pi ( n + \gamma)$ with~:
\begin{eqnarray}
\label{gamlow}
\gamma_l^{S}(E,x_c)=\frac{1}{\pi}\arctan{ X_l^{S}} \\
\gamma_l^{AS}(E,x_c)=\frac{1}{\pi}\arctan{ X_l^{AS}}+1
\end{eqnarray}
Since the energy $E$ is simply related to the action $E=1/2+S_l/\pi$, we finally obtain the two implicit equations~:
\begin{eqnarray}
\label{eqimplls}
&E_n^{S}=n + \frac{1}{2} + \frac{1}{\pi}\arctan{X_l^{S}}\\
\label{eqimpllas}
&E_n^{AS}=n+\frac{3}{2}-\frac{1}{\pi}\arctan{X_l^{AS}}
\end{eqnarray}
from which we obtain the energy levels $E_n(x_c)$ in the region $R_l \leq R_r\leq \left|x_c\right|$.

																				\subsection{Region $\cal D$~: $ R_l \leq \left|x_c\right|\leq R_r$}
																							
The region ${\cal D}$ represented in Fig.  \ref{fig:potential2}   does not exist for the symmetric potential $V_{Sym}(x)$. We have now to  linearize the  potential around $-\left|x_c\right|+R_l$ for the left well and around $x=0$ for the right well. The solution inside the left well takes the familiar form~:
\begin{eqnarray}
\Psi^l(x)&=&\bar{\alpha_l} \Ai(a_lx+x_l)+\bar{\beta_l} \Bi(a_lx+x_l)\\
a_l^3&=&2 R_l\\
x_l&=&(2R_l)^{1/3}\left(R_l-\left|x_c\right|\right)
\end{eqnarray}
whereas the one inside the right well is~:
\begin{eqnarray}
\Psi^r(x)&=&\bar{\alpha_r} \Ai(y_r-a_r x)+\bar{\beta_r} \Bi(y_r-a_r x)\\
a_r^3&=& \left|2x_c\right|\\
y_r&=&\frac{x_c^2-R_r^2}{\left|2x_c\right|^{2/3}}.
\end{eqnarray}
The next step consists  in the matching of the Airy functions with the WKB approximation valuable inside a well.	The treatment for the left well has been performed in the paragraph \ref{sec:symwellle}, of this appendix and gives:		
\begin{eqnarray}
\tan S_l =\frac{\alpha_l}{\beta_l}=\frac{\bar{\alpha_l}}{\bar{\beta_l}}
\label{alphalie}
\end{eqnarray}	
whereas the calculation for the right well has been made in the paragraph \ref{sec:symwellhe} and gives~: 
\begin{eqnarray}
\label{alpharie}
\tan( S_r+ \delta_r) &=& \frac{\alpha_r}{\beta_r}=\frac{\bar{\alpha_r}}{\bar{\beta_r}} \\
\text{where}\ \ \ \ \ \ \
\delta_r &=& \frac{\left(R_r^2-x_c^2\right)^{3/2}}{3\left|x_c\right|} \ .
\end{eqnarray}																			
The current conservation and the continuity of the wave function at $x=0$ gives the two equations~:
\begin{eqnarray*}
\bar{\alpha}_l \Ai(x_l)+ \bar{\beta}_l \Bi(x_l) &=& \bar{\alpha}_r \Ai(y_r)+ \bar{\beta}_r \Bi(y_r)\\
\bar{\alpha_l} a_l \Ai'(x_l)+ \bar{\beta_l} a_l \Bi'(x_l) &=& -\bar{\alpha}_r a_r \Ai'(y_r)- \bar{\beta}_r a_r \Bi'(y_r)
\end{eqnarray*}
from where we extract the ratio $\frac{\bar{\alpha}_r}{\bar{\beta}_r}$:
\begin{widetext}
\begin{equation}
\frac{\bar{\alpha}_r}{\bar{\beta}_r}=-\frac{\frac{\bar{\alpha}_l}{\bar{\beta}_l}\left( \Bi'(y_r)  \Ai(x_l) + \left(\frac{R_l}{\left|x_c\right|}\right)^{1/3}\Bi(y_r)\Ai'(x_l) \right) +\Bi'(y_r)\Bi(x_l)+ \left(\frac{R_l}{\left|x_c\right|}\right)^{1/3}\Bi(y_r) \Bi'(x_l)}{\frac{\bar{\alpha}_l}{\bar{\beta}_l}\left(\Ai'(y_r)   \Ai(x_l)+\left(\frac{R_l}{\left|x_c\right|}\right)^{1/3} \Ai(y_r)\Ai'(x_l) \right)+ \Ai'(y_r)\Bi(x_l)+\left(\frac{R_l}{\left|x_c\right|}\right)^{1/3} \Ai(y_r)\Bi'(x_l)}
\label{alphabetaas2}
\end{equation}
Introducing the total action $S_t=S_l+S_r$ and  using (\ref{alphalie}) and (\ref{alpharie}) we obtain the relation~:
\begin{eqnarray}
\frac{\alpha_r}{\beta_r}=\frac{\tan(S_t+\delta_r)-\frac{\alpha_l}{\beta_l}}{1+\tan(S_t+\delta_r)\frac{\alpha_l}{\beta_l}}=\frac{\nu-X_l}{1+\nu X_l},
\end{eqnarray}
where we have introduced $\nu\equiv\tan{(S_t+\delta_r)}$. We inject this last relation into (\ref{alphabetaas2})  to obtain again a polynomial in $X_l$~:
\begin{eqnarray}
\label{polas2}
\left\{ \nu \left(\Bi'(y_r)\Ai(x_l)+\left(\frac{R_l}{\left|x_c\right|}\right)^{1/3}\Bi(y_r)\Ai'(x_l)\right)-\Ai'(y_r)\Ai(x_l)- \left(\frac{R_l}{\left|x_c\right|}\right)^{1/3}\Ai'(x_l) \Ai(y_r)\right\}X_l^2\\
+\left\{\nu \left( \Ai'(y_r)\Ai(x_l)+\Bi'(y_r)\Bi(x_l)+ \left(\frac{R_l}{\left|x_c\right|}\right)^{1/3}\left[\Ai'(x_l)\Ai(y_r)+\Bi'(x_l)\Bi(y_r)\right]\right)
\right.
\nonumber\\
\left.
+  \Bi'(y_r)\Ai(x_l)-\Ai'(y_r)\Bi(x_l)+ \left(\frac{R_l}{\left|x_c\right|}\right)^{1/3}\left[\Bi(y_r)\Ai'(x_l)-\Ai(y_r)\Bi'(x_l) \right] \right\}X_l\nonumber\\
+\nu\left(\Ai'(y_r)\Bi(x_l)+\left(\frac{R_l}{\left|x_c\right|}\right)^{1/3}\Ai(y_r)\Bi'(x_l)\right)+\Bi'(y_r)\Bi(x_l)+\left(\frac{R_l}{\left|x_c\right|}\right)^{1/3}\Bi(y_r)\Bi'(x_l)=0\nonumber .
\end{eqnarray}
\end{widetext}
Note that this polynomial satisfies the continuity of the spectrum on the parabola for which $x_c=R_r$. Indeed, in this case we have $\delta_r=0$ and $x_r=y_r=0$ so that the two polynomials (\ref{polas1}) in region $\cal C$ and (\ref{polas2}) in region $\cal D$ coincide and give the same solutions. From here, the work is exactly the same than for the lower region. Therefore, we have just to take the two solutions $X_l^S$ and $X_l^{AS}$ of this polynomial and insert them into (\ref{gamlow}) to obtain the left MI in the intermediate region. Then we insert them into (\ref{eqimplls}) and (\ref{eqimpllas}) to obtain the implicit equations in $E_n$ and $x_c$ from where we can extract the spectrum $E_n(x_c)$ in the region $\cal D$, characterized by the relation $R_l \leq \left|x_c\right|  \leq R_r$.

																				\subsection{Region $\cal E$~: $ \left|x_c\right|\leq  R_l\leq R_r$}
The last region ${\cal E}$ satisfies $\left|x_c\right|\leq  R_l\leq R_r$ and is illustrated in Fig.  \ref{fig:potential2}. As for the high energy region of the double symmetric well, we decompose the well into two parts from $x=0$, and keep working with $S_l$ that is the action in the well where $x<0$. The strategy is still the same, and the calculations are very similar to those in section \ref{sec:symwellhe}. We express the ratio $\alpha/\beta$ from two ways. First, we obtain a relation of the type (\ref{slhe}) thanks to the matching of the WKB wave function with the solutions of the linearised Schr\"odinger equation around $x=0$ that we express as~:
\begin{align*}
&\Psi^l(x)=\bar{\alpha_l} \Ai(a_lx+y_l)+\bar{\beta_l} \Bi(a_lx+y_l)\ \ \ \text{for}\  x<0\\
&\Psi^r(x)=\bar{\alpha_r} \Ai(y_r-a_rx)+\bar{\beta_r} \Bi(y_r-a_rx)\ \ \ \text{for}\  x>0\\
& \hspace{1.5cm} y_l=\frac{x_c^2-R_l^2}{\left|2x_c\right|^{2/3}}\hspace{1cm} y_r=\frac{x_c^2-R_r^2}{\left|2x_c\right|^{2/3}}\\
& \hspace{3cm} a_l^3=a_r^3=\left|2x_c\right| \ .
\end{align*}
The difference here comes from the asymmetry $V_0=1$ so that the relation (\ref{slhe}) becomes~:
\begin{eqnarray}
\label{tansle}
&&\tan{(S_l+\delta_l)}=  X_l=\frac{\alpha_l}{\beta_l} \\
&&\tan{(S_r+\delta_r)}=   X_r=\frac{\alpha_r}{\beta_r} \\
&&\delta_r = \frac{\left(R_r^2-x_c^2\right)^{3/2}}{3\left|x_c\right|}\ \ \ \ \delta_l = \frac{\left(R_l^2-x_c^2\right)^{3/2}}{3\left|x_c\right|}
\end{eqnarray}
In order to obtain a polynomial  in terms of $X_l$ in the high energy region, we use the lattest relations to express $X_r$ as a function of $X_l$~:
\begin{eqnarray}
X_r&=&\frac{\tan{(S_t+\delta_r)-\tan{S_l}}}{1+\tan{S_l}\tan{(S_t+\delta_r)}}\notag\\
&=&\frac{(\nu \tan{\delta_l}-1)\frac{\alpha_l}{\beta_l}+\nu+\tan{\delta_l}}{(\nu+\tan{\delta_l})\frac{\alpha_l}{\beta_l}-(\nu \tan{\delta_l-1})}\notag\\
X_r&=&\frac{\sigma \ X_l+\tau}{\tau \ X_l -\sigma}
\label{xrhe}
\end{eqnarray}
where we have introduced $\sigma\equiv\nu \tan{\delta_l}-1$ and $\tau\equiv\nu+\tan{\delta_l}$. Then, we use the continuity conditions at $x=0$ to obtain~:
\begin{widetext}
\begin{eqnarray}
X_r=\frac{\bar{\alpha}_r}{\bar{\beta}_r}=-\frac{X_l\left[ \Bi'(y_r)  \Ai(y_l) + \Bi(y_r)\Ai'(y_l) \right] +\Bi'(y_r)\Bi(y_l)+ \Bi(y_r) \Bi'(y_l)}{X_l\left[\Ai'(y_r)   \Ai(y_l)+ \Ai(y_r)\Ai'(y_l) \right]+ \Ai'(y_r)\Bi(y_l)+ \Ai(y_r)\Bi'(y_l)}
\label{alphabetaas3}
\end{eqnarray}
>From Eqs. (\ref{xrhe}) and (\ref{alphabetaas3}), we  obtain a polynomial  for $X_l$~:
\begin{eqnarray}
\label{polas3}
\left\{ \tau \left(\Bi'(y_r)\Ai(y_l)+\Bi(y_r)\Ai'(y_l)\right)+\sigma\left(\Ai'(y_r)\Ai(y_l)+  \Ai'(y_l)\Ai(y_r)\right)\right\}X_l^2\\
+\left\{\tau \left( \Ai'(y_r)\Ai(y_l)+\Bi'(y_r)\Bi(y_l)+\Ai'(y_l)\Ai(y_r)+ \Bi'(y_l)\Bi(y_r)\right)
\right. \notag\\ \left.
- \sigma \left(\Bi'(y_r)\Ai(y_l)-\Ai'(y_r)\Bi(y_l)+\Bi(y_r)\Ai'(y_l)-\Ai(y_r)\Bi'(y_l)\right)\right\}X_l\notag\\
+\tau \left(\Ai'(y_r)\Bi(y_l)+\Ai(y_r)\Bi'(y_l)\right)-\sigma \left(\Bi'(y_r)\Bi(y_l)+\Bi(y_r)\Bi'(y_l)\right)=0 \notag
\end{eqnarray}
\end{widetext}
Again, it is easy to check that for $\left|x_c\right|=R_l$, we have $\tau=\nu$, $\sigma=-1$ and $x_l=y_l=0$ so that the polynomials (\ref{sec:symwellhe}) in region $\cal D$  and (\ref{polas3}) in region $\cal E$  coincide, what assures the continuity of the spectrum between these two regions. To obtain the implicit equations and extract the spectrum $E_n(x_c)$ in this region, we inject the solutions $X^{S/AS}$ into (\ref{tansle}) to obtain an expression of the mismatch indexs $\gamma^{S/AS}$ like in  region $\cal B$ (see Eq. (\ref{mile2}) with $\delta\rightarrow\delta_l$) and write explicitly the action $S_l$ as a function of the energy and $x_c$ as in the expression (\ref{action2}) with $R\rightarrow R_l$. 


\end{document}